\documentclass[11pt, a4paper]{article}
\usepackage{jheparxiv}
\usepackage[utf8]{inputenc}
\usepackage{amsmath}
\usepackage{amsfonts}
\usepackage{amssymb}
\usepackage{latexsym}
\usepackage{mathrsfs}
\usepackage{braket}		
\usepackage{graphicx}
\usepackage{color}
\usepackage{xcolor}
\usepackage{slashed}
\usepackage{twistor}
\usepackage[all]{xy}

\newcommand{\tpsi}{\tilde{\psi}}
\newcommand{\sa}{\mathsf{a}}

\newcommand{\rg}{\mathrm{g}}
\newcommand{\DZ}{\int\text{D}^{3|4}Z\wedge}

\newcommand{\sT}{\mathsf{T}}

\renewcommand{\d}{\mathrm{d}}

\newcommand{\qed}{\hfill \ensuremath{\Box}}

\subheader{\hfill \texttt{IMPERIAL-TP-TA-2019-03}}

\title{Twistor fishnets}

\author{Tim Adamo \& Sumer Jaitly}

\affiliation{Theoretical Physics Group, Blackett Laboratory \\
        Imperial College London, SW7 2AZ, United Kingdom}

\emailAdd{t.adamo@imperial.ac.uk}
\emailAdd{sumer.jaitly14@imperial.ac.uk}

\abstract{Four-dimensional conformal fishnet theory is an integrable scalar theory which arises as a double scaling limit of $\gamma$-deformed maximally supersymmetric Yang-Mills. We give a perturbative reformulation of $\gamma$-deformed super-Yang-Mills theory in twistor space, and implement the double scaling limit to obtain a twistor description of conformal fishnet theory. The conformal fishnet theory retains an abelian gauge symmetry on twistor space which is absent in space-time, allowing us to obtain cohomological formulae for scattering amplitudes that manifest conformal invariance. We study various classes of scattering amplitudes in twistor space with this formalism.}

\begin{document}

\maketitle
 
\section{Introduction}

Over the last two decades, there have been many remarkable advances in the study of observables such as scattering amplitudes and correlation functions in a wide variety of quantum field theories (QFTs). In many cases, an underlying motivation has been that -- at least for some QFTs -- it might be possible to find alternative formulations which manifest hidden symmetries, streamline perturbative calculations or enable the \emph{exact} computation of observables. One of the most widely studied QFTs in this context has been planar, maximally supersymmetric Yang-Mills theory in four-dimensions ($\cN=4$ SYM). 

Indeed, planar $\cN=4$ SYM has many advantageous properties: it is exactly conformal, has only a single tuneable parameter (the 't Hooft coupling), admits a holographic description, possesses an infinite-dimensional symmetry algebra and is widely believed to be an integrable conformal field theory (CFT) (cf., \cite{Beisert:2010jr,Elvang:2013cua}). Yet there are issues (e.g., IR-divergences) which obstruct the analytic determination of observables in this theory.

Recently, there has been considerable interest in other QFTs for which it may be possible to make more exact analytic statements. A prime example is the four-dimensional \emph{conformal fishnet theory} (FCFT)~\cite{Gurdogan:2015csr}, whose name derives from the shape of its typical Feynman diagrams in the planar limit. \emph{A priori}, this FCFT seems far removed from $\cN=4$ SYM. It is a non-unitary theory of two complex scalars valued in the adjoint representation of SU$(N)$ with only quartic interactions; there is no supersymmetry or local gauge symmetry. However, FCFT can actually be derived by taking a `double scaling limit' of the well-known $\gamma$-deformation of $\cN=4$ SYM~\cite{Leigh:1995ep,Lunin:2005jy,Frolov:2005dj,Frolov:2005ty}. The double scaling limit decouples the gauge field, leaving a chiral field theory of fermions and scalars ($\chi$FT)~\cite{Gurdogan:2015csr,Caetano:2016ydc,Kazakov:2018gcy}. Setting two of the couplings in this $\chi$FT to zero leaves only two complex scalars with a chiral quartic interaction. 

Quantum completeness requires FCFT to be supplemented with quartic double trace interactions, whose couplings have non-trivial $\beta$-functions~\cite{Fokken:2013aea,Sieg:2016vap}. This full theory has two conformal fixed points, which have been determined perturbatively up to seven loops~\cite{Sieg:2016vap,Grabner:2017pgm}. At these conformal fixed points and in the planar limit, FCFT is believed to be integrable\footnote{The integrability of fishnet diagrams was already established almost forty years ago~\cite{Zamolodchikov:1980mb}.}~\cite{Chicherin:2017cns,Gromov:2017cja,Chicherin:2017frs,Grabner:2017pgm,Kazakov:2018hrh}. 

Anomalous dimensions, certain correlation functions and scattering amplitudes have all been determined \emph{exactly} as functions of the coupling in FCFT~\cite{Grabner:2017pgm,Gromov:2018hut,Korchemsky:2018hnb,Basso:2018cvy,Chowdhury:2019hns}. In addition, FCFT appears to admit a holographic description in terms of a discretized string-like model~\cite{Basso:2018agi,Gromov:2019aku,Gromov:2019bsj,Gromov:2019jfh}, and provides a first example of a non-supersymmetric 4d QFT with spontaneously broken conformal invariance~\cite{Karananas:2019aab}. Fishnet CFTs with the hallmarks of integrability also exist in diverse dimensions~\cite{Kazakov:2018qez,Derkachov:2018rot} and arise from non-integrable gauge theories~\cite{Pittelli:2019ceq}. Furthermore, many of the Feynman integrals of FCFT play an important role in other integrable QFTs like $\cN=4$ SYM (cf., \cite{Basso:2017jwq,Bourjaily:2018ycu}). 

\medskip

One of many `non-standard' tools used in modern studies of $\cN=4$ SYM is \emph{twistor theory}~\cite{Penrose:1967wn} (cf., \cite{Penrose:1972ia,Ward:1990vs,Adamo:2017qyl} for reviews). This began with the realization that the full tree-level S-matrix of $\cN=4$ SYM is given by genus zero worldsheet correlation functions of a string theory whose target is twistor space~\cite{Witten:2003nn,Roiban:2004yf}, but there are myriad other applications of twistor theory in this setting. For instance, $\cN=4$ SYM can be described perturbatively by a \emph{twistor action} functional~\cite{Mason:2005zm,Boels:2006ir}, whose gauge freedom is significantly greater than the standard space-time gauge transformations. This, combined with the natural manifestation of superconformal symmetry and the non-local relation between space-time and twistor space, makes the twistor action a very useful tool for the study of perturbative $\cN=4$ SYM (cf., \cite{Adamo:2011pv})\footnote{The twistor approach to $\cN=4$ SYM is equivalent to that of Lorentz harmonic chiral superspace~\cite{Sokatchev:1995nj,Chicherin:2016soh}; this paper will be phrased entirely in the language of twistors.}. 

The twistor action provides a natural derivation of modified Feynman rules (i.e., the MHV formalism~\cite{Cachazo:2004kj}) for $\cN=4$ SYM~\cite{Boels:2007qn,Bullimore:2010pj,Adamo:2011cb}, can be used to provide descriptions of local operators and form factors in twistor space~\cite{Chicherin:2014uca,Chicherin:2016fac,Chicherin:2016fbj,Koster:2016ebi,Koster:2016loo,Chicherin:2016qsf,Koster:2016fna,Brandhuber:2016xue}, and characterizes dual superconformal symmetry~\cite{Mason:2009qx} and its breaking~\cite{Bullimore:2011kg}. It gives a powerful method for proving various dualities and correspondences in planar $\cN=4$ SYM to all loop orders (at the level of the loop integrand~\cite{ArkaniHamed:2010kv}) which were not possible with space-time methods. These include the scattering amplitude/Wilson loop duality~\cite{Mason:2010yk,Bullimore:2011ni}, as well as correspondences between null limits of correlations functions and null polygonal Wilson loops~\cite{Adamo:2011dq,Adamo:2011cd}. 

Twistor actions have been developed for many other QFTs, including self-dual Einstein (super)gravity~\cite{Mason:2007ct}, four-dimensional conformal (super)gravity~\cite{Mason:2005zm,Adamo:2013tja}, conformal higher spin theory~\cite{Haehnel:2016mlb,Adamo:2016ple} and three-dimensional (super-)Yang-Mills-Higgs theory~\cite{Adamo:2017xaf}. Generally, one expects twistor methods to be useful for any four-dimensional massless QFT, especially if there is conformal invariance. It is thus natural to consider a twistor formulation of FCFT. 

\medskip

In this paper, we give twistor actions for $\gamma$-deformed $\cN=4$ SYM, $\chi$FT and FCFT, showing how to implement the $\gamma$-deformation and double scaling limit on twistor space. For each of these theories, the twistor action has an enhanced gauge invariance relative to space-time. Focusing on FCFT, we impose a twistor axial gauge, derive Feynman rules suitable for computing scattering amplitudes in a `cohomological' representation which is natural in twistor space~\cite{Adamo:2011cb,Adamo:2013cra} and use them to characterize the appearance of UV-divergences and double trace counterterms. We then compute twistor formulae (which manifest conformal invariance) for various scattering amplitudes in the planar limit and at one of the conformal fixed points.

Twistor amplitude formulae previously obtained for $\cN=4$ SYM had the drawback of only being well-defined at the integrand level. This is due to IR-divergences and the lack of an easy-to-use regulator on twistor space, though some concrete calculations have been possible using mass regularization~\cite{Lipstein:2013xra}. The absence of IR- and UV-divergences in FCFT ensures that our formulae define meaningful amplitudes.

Of course, an interesting long-term goal of this research is to provide a twistorial description of the integrability underlying FCFT. This aim has already been explored for planar $\cN=4$ SYM with limited success~\cite{Koster:2014fva}, but the added simplicity of FCFT -- and the fact that the twistor description is \emph{exact} rather than perturbative -- gives cause to be optimistic. This paper sets the foundations and assembles the toolkit needed to embark on the study of FCFT integrability in twistor space in future work.

\medskip

The paper is organized as follows: section~\ref{GDEF} begins with a brief review of twistor theory and the twistor action for $\cN=4$ SYM, then develops the $\gamma$-deformation of $\cN=4$ SYM on twistor space and proves its perturbative equivalence to the space-time theory. In section~\ref{DSLIM}, we implement the double scaling limit on twistor space, and prove that the resulting twistor action describes $\chi$FT; decoupling all fermions results in a quartic theory on twistor space which we prove to be equivalent to classical FCFT. 

Section~\ref{FRULES} reviews the cohomological representation of scattering amplitudes in twistor space and derives the Feynman rules for the FCFT twistor action in axial gauge. We provide a characterization of UV-divergences on twistor space in terms of certain structures appearing in cohomological amplitudes, and complete the twistor action with the double trace counterterms needed to remove these divergences. In section~\ref{SCAMPS} we compute the cohomological amplitudes for various scattering processes in FCFT, including a class of tree-level exact amplitudes and general fishnet diagrams.


\section{Twistor action for $\gamma$-deformed $\cN=4$ SYM}
\label{GDEF}

While it is straightforward to write the twistor action of conformal fishnet theory (FCFT) directly from its space-time Lagrangian, it is instructive to derive the theory on twistor space itself. In this section, we give a perturbative description of $\gamma$-deformed $\cN=4$ super-Yang-Mills (SYM) theory in twistor space, proceeding from the well-known twistor action of $\cN=4$ SYM~\cite{Mason:2005zm,Boels:2006ir,Adamo:2011pv}. After a brief review of the basics of twistor geometry and the $\cN=4$ SYM twistor action, we implement the $\gamma$-deformation on twistor space through a $\star$-product which acts on the anti-commuting twistor variables and prove that the resulting twistor action is perturbatively equivalent to $\gamma$-deformed $\cN=4$ SYM.


\subsection{Twistor theory and $\cN=4$ SYM}

The twistor space $\PT$ appropriate to four-dimensional $\cN=4$ supersymmetry is an open subset of the complex projective supermanifold $\P^{3|4}$. This twistor space is charted with homogeneous coordinates $Z^I$, with the index $I$ ranging over four commuting (bosonic) and four anti-commuting (fermionic) directions:
\be\label{4PTcoord}
Z^I=(Z^{A},\, \chi^{a})=(\mu^{\dot\alpha},\,\lambda_{\alpha},\,\chi^{a})\,,
\ee
where $A$ is a SL$(4,\C)$ spinor index and $\mu^{\dot\alpha}$, $\lambda_{\alpha}$ are commuting SL$(2,\C)$ spinors of opposite chirality. The coordinates $\chi^{a}$ are anti-commuting, with the index $a=1,\ldots,4$. As homogeneous coordinates, $Z^I$ are defined only up to an overall $\C^*$ rescaling, with $Z^I\sim rZ^I$ for $r$ any non-zero complex number.

The twistor space $\PT$ is related non-locally to space-time through the `incidence relations,' which have a simple algebraic expression:
\be\label{4PTinc}
\mu^{\dot\alpha}=x^{\alpha\dot\alpha}\lambda_{\alpha}\,, \qquad \chi^{a}=\theta^{\alpha a}\lambda_{\alpha}\,,
\ee
where $(x^{\alpha\dot\alpha},\theta^{\alpha a})$ are coordinates on complexified, chiral $\cN=4$ Minkowski superspace. These incidence relations indicate that every point $(x,\theta)\in\M$ corresponds to a holomorphic, linearly embedded Riemann sphere -- or `line' -- in twistor space, $X\cong\P^1\subset\PT$. The conformal structure of $\M$ is encoded by these twistor lines: two twistor lines $X,Y$ intersect if and only if their corresponding space-time points $(x,\theta)$, $(y,\theta')$ are null separated. 

Since $\mathfrak{psl}(4|4,\C)$ acts projectively on $\P^{3|4}$, twistor space carries a natural linear action of the complexified superconformal algebra. Conformal invariants are naturally built from the SL$(4,\C)$-invariant Levi-Civita symbol $\epsilon_{ABCD}$. For instance, any four points in $\PT$ define the conformal invariant
\be\label{TConinv}
(ijkl):=\epsilon_{ABCD}\,Z^{A}_{i}\,Z^{B}_{j}\,Z^{C}_{k}\,Z^{D}_{l}\,,
\ee
and more general superconformal invariants are also easily constructed (cf., \cite{Mason:2009qx}).

There are two main theorems of twistor theory which will be of particular importance. The first is the \emph{Penrose transform}, which states that free zero-rest-mass fields on $\M$ can be described by cohomology classes on $\PT$~\cite{Penrose:1969ae,Eastwood:1981jy}:
\be\label{PenTran}
\left\{\mbox{Free z.r.m. fields on } \M \mbox{ of helicity } h\right\}\cong H^{1}(\PT,\,\cO(2h-2))\,,
\ee
where $\cO(k)$ is the sheaf of holomorphic functions, homogeneous of weight $k\in\Z$ on $\PT$. The second is the \emph{Ward correspondence}~\cite{Ward:1977ta}, which relates self-dual Yang-Mills connections on $\M$ to holomorphic vector bundles on $\PT$ which obey some technical conditions\footnote{For gauge group SU$(N)$, every Yang-Mills instanton on $\M$ corresponds to a rank $N$ holomorphic vector bundle which is trivial on restriction to every line in $\PT$, admits a positive real form, and has a trivial determinant line bundle (cf., \cite{Ward:1990vs}).}.

To pass from complexified Minkowski space $\M$ to a real space-time, reality conditions must be imposed on twistor space. We will primarily use Euclidean reality conditions, associated with the real slice $\R^{4|8}\subset\M$. In this case, the reality structure is encoded by a `quaternionic' conjugation operation, defined on Weyl 2-spinors and the fermionic twistor variables by~\cite{Woodhouse:1985id}:
\be\label{euc1}
\mu^{\dot\alpha}=(\mu^{\dot0},\,\mu^{\dot{1}})\mapsto \hat{\mu}^{\dot\alpha}=\left(-\overline{\mu^{\dot1}},\, \overline{\mu^{\dot0}}\right)\,, 
\ee
\begin{equation*}
\chi^{a}=(\chi^1,\,\chi^2,\,\chi^3,\,\chi^4)\mapsto \hat{\chi}^a=(-\overline{\chi^2},\,\overline{\chi^1},\,-\overline{\chi^4},\,\overline{\chi^3})\,,
\end{equation*}
and similarly for un-dotted spinors. This induces an involution $\sigma:\PT\rightarrow\PT$ via $Z^I\mapsto \hat{Z}^{I}$, and from \eqref{euc1} it is easy to see that $\sigma^{2}=-\mathrm{id}$ on $\PT$. There are no real points under $\sigma$, but lines obeying $X\cong \hat{X}$ \emph{are} preserved and correspond to points in $\R^{4|8}$.

So for Euclidean reality conditions, twistor space fibres over $\P^1$:
\begin{equation*}
 \xymatrix{
 \P^1 \ar@{^{(}->}[r] & \PT\cong\R^{4|8}\times\P^1 \ar[d] \\
  & \R^{4|8}}
\end{equation*}
The incidence relations take points in $\R^{4|8}$ to twistor lines as before, but now each point in $\PT$ corresponds to a unique point in $\R^{4|8}$ via:
\be\label{euc2}
x^{\alpha\dot\alpha}=\frac{\hat{\mu}^{\dot\alpha}\lambda^{\alpha}-\mu^{\dot\alpha}\hat{\lambda}^{\alpha}}{\la\lambda\,\hat{\lambda}\ra}\,, \qquad \theta^{\alpha a}=\frac{\hat{\chi}^a \lambda^{\alpha}-\chi^{a}\hat{\lambda}^{\alpha}}{\la\lambda\,\hat{\lambda}\ra}\,,
\ee
with $\la\lambda\,\hat{\lambda}\ra=\lambda^{\alpha}\hat{\lambda}_{\alpha}=\epsilon^{\alpha\beta}\lambda_{\beta}\hat{\lambda}_{\alpha}$. This means that the homogeneous coordinates $Z^I$ can be interchanged with $(x,\theta,\lambda)$ as coordinates on $\PT$, with the projective scaling carried exclusively by $[\lambda_\alpha]$, now viewed as homogeneous coordinates on the $\P^1$ fibres. For instance, a useful basis of the anti-holomorphic tangent and cotangent bundles of $\PT$ can be given in these coordinates~\cite{Woodhouse:1985id}:
\be\label{eucb1}
\dbar_0=\la\lambda\,\hat{\lambda}\ra\,\lambda^{\alpha}\frac{\partial}{\partial\hat{\lambda}^\alpha}\,, \qquad \dbar_{\dot\alpha}=\lambda^{\alpha}\frac{\partial}{\partial x^{\alpha\dot\alpha}}\,, 
\ee
\be\label{eucb2}
\bar{e}^0=\frac{\la\hat{\lambda}\,\d\hat{\lambda}\ra}{\la\lambda\,\hat{\lambda}\ra^2}\,, \qquad \bar{e}^{\dot\alpha}=\frac{\hat{\lambda}_{\alpha}\,\d x^{\alpha\dot\alpha}}{\la\lambda\,\hat{\lambda}\ra}\,.
\ee
Using the incidence relations, one confirms that $\bar{e}^0\dbar_0+\bar{e}^{\dot\alpha}\dbar_{\dot\alpha} = \d\hat{Z}^{A}\frac{\partial}{\partial \hat{Z}^A}$, as appropriate for the complex structure on $\PT$ with these reality conditions.

\medskip

With these tools, a perturbative description of $\cN=4$ SYM can be given entirely on twistor space. Since Yang-Mills theory admits a perturbative expansion around the self-dual sector~\cite{Chalmers:1996rq}, this is achieved using the Ward correspondence to give a twistorial description of the instanton sector and the Penrose transform to describe perturbations away from self-duality~\cite{Mason:2005zm,Boels:2006ir}. The resulting twistor action is a functional of a single field $\cA\in\Omega^{0,1}(\PT,\cO)$ valued in the adjoint representation of the gauge group. By expanding in the fermionic directions of twistor space,
\be\label{4TF}
\cA=a+\chi^{a}\,\tilde{\psi}_{a}+\frac{\chi^{a}\chi^{b}}{2}\,\phi_{ab}+\frac{1}{3!}\epsilon_{abcd}\chi^{a}\chi^{b}\chi^{c}\,\psi^{d}+\frac{\chi^4}{4!}\,b\,,
\ee
where each remaining component is a $(0,1)$-form of the appropriate homogeneity on the bosonic twistor space. When $\dbar \cA=0$, the Penrose transform states that $\cA$ contains the full linear spectrum of $\cN=4$ SYM.

The twistor action for $\cN=4$ SYM is given by~\cite{Boels:2006ir}
\be\label{4TA1}
S[\cA]=\frac{\im}{2\,\pi}\int_{\PT}\!\D^{3|4}Z\wedge\tr\left(\cA\wedge\dbar\cA+\frac{2}{3}\cA\wedge\cA\wedge\cA\right)+\rg^{2}\oint_{\R^{4|8}} \!\d^{4|8}X\,\mathrm{log}\,\mathrm{det}\!\left(\dbar+\cA\right)|_{X}\,,
\ee
where $\rg$ is the dimensionless coupling constant. The first term in this action is local on $\PT$, with $\D^{3|4}Z$ standing for the global holomorphic section of the canonical bundle
\be\label{Ber1}
\D^{3|4}Z=\D^{3}Z\,\d^{4}\chi=\epsilon_{ABCD}\,Z^{A}\,\d Z^{B}\wedge\d Z^{C}\wedge\d Z^{D}\,\d^{4}\chi\,.
\ee
The second, non-local, contribution to the action is integrated over the real contour in the moduli space of lines in $\PT$ corresponding to $\R^{4|8}$, and the integrand can be understood perturbatively via the expansion
\begin{multline}\label{logdet}
\mathrm{log}\,\mathrm{det}\!\left(\dbar+\cA\right)|_{X}=\tr\left(\log \dbar|_X\right) \\
+\sum_{n=2}^{\infty}\frac{1}{n}\left(\frac{1}{2\pi\im}\right)^{n} \int\limits_{(\P^1)^{n}} \frac{\D\sigma_{1}\cdots\D\sigma_{n}}{(1\,2)\,(2\,3)\cdots (n\,1)}\,\tr\left(\cA_1\,\cA_2\cdots \cA_n\right)\,.
\end{multline}
Here, $\sigma_{i}^{\alpha}=(\sigma_{i}^{0},\,\sigma_{i}^{1})$ are homogeneous coordinates on the $i^{\mathrm{th}}$ copy of the twistor line $X\cong\P^1$; $(i\,j):=\sigma_{i}^{\alpha}\sigma_{j\,\alpha}$ is the SL$(2,\C)$-invariant inner product on these coordinates; and $\D\sigma_i:=(\sigma_i\,\d\sigma_{i})$ is shorthand for the weight $+2$ holomorphic measure on the $i^{\mathrm{th}}$ copy of $\P^1$. Similarly, $\cA_i$ is shorthand for $\cA(Z(\sigma_i))$, the twistor field of \eqref{4TF} evaluated at the point $\sigma_i$ on the twistor line $X$.

The action \eqref{4TA1} is invariant with respect to non-abelian gauge transformations on twistor space,
\be\label{4gi}
\dbar+\cA\rightarrow g\left(\dbar+\cA\right) g^{-1}\,,
\ee
for $g$ any weight zero function on $\PT$ valued in the adjoint of the gauge group. This gauge freedom is much greater than the usual space-time gauge invariance, since space-time gauge transformations are functions of only four bosonic variables. Nevertheless, the twistor action \eqref{4TA1} is perturbatively equivalent to the space-time Lagrangian of $\cN=4$ SYM. Indeed, for a particular choice of gauge on $\PT$, the twistor action is equal to the Chalmers-Siegel Lagrangian of $\cN=4$ SYM on $\R^{4}$~\cite{Boels:2006ir,Koster:2017fvf}.


\subsection{The $\gamma$-deformed twistor action}

The $\gamma$-deformation of $\cN=4$ SYM breaks the PSL$(4|4,\C)$ superconformal group to SL$(4,\C)$ $\times$U$(1)^{3}$; the three deformation parameters $\{\gamma_i\}_{i=1,2,3}$ correspond to the Cartan subgroup U$(1)^3$ of the initial SO$(6)$ R-symmetry group. On space-time, this deformation can be operationalized through a non-commutative $\star$-product on the field space of $\cN=4$ SYM~\cite{Lunin:2005jy,Fokken:2013aea}, which introduces phase factors depending on the U$(1)^3$ charges of the fields. On twistor space, the $\gamma$-deformation is most naturally described by a $\star$-product on polynomials of the fermionic twistor coordinates~\cite{Kulaxizi:2004pa,Gao:2006mw}.

We assign a U$(1)^3$ charge vector $q^k$ to every power of the $\chi^{a}$ twistor variables. Using the notation $\chi^{a}=(\chi^i,\chi^4)$, for $i=1,2,3$, the charge assignments are:
\be\begin{split}\label{qcharge}
   q^{k}[1]=0\,, & \qquad  q^{k}[(\chi)^4]=0\,, \\
   q^{k}[\chi^i]=\left\{\begin{array}{cc}
                         -\frac{1}{2} & \mbox{if } k=i \\
                         \frac{1}{2} & \mbox{otherwise}
                        \end{array}\right.
\,, & \qquad q^{k}[\chi^4]=\left(-\frac{1}{2},\,-\frac{1}{2},\,-\frac{1}{2}\right)\,, \\
   q^{k}[\chi^i \chi^4]=-\delta^{ki}\,, & \qquad q^{k}[\chi^{i}\chi^{j}]=|\epsilon^{kij}|\,, \\
   q^{k}[(\chi)^3_{i}]=\left\{\begin{array}{cc}
                         \frac{1}{2} & \mbox{if } k=i \\
                         -\frac{1}{2} & \mbox{otherwise}
                        \end{array}\right.\,, & \qquad q^{k}[(\chi)^3_{4}]=\left(\frac{1}{2},\,\frac{1}{2},\,\frac{1}{2}\right)\,,
   \end{split}
\ee
where $(\chi)^{3}_{a}:=\epsilon_{bcda}\chi^{b}\chi^{c}\chi^{d}$. These charges define a $\star$-product on the fermionic twistor coordinates. Let $A$ and $B$ be any two powers of the $\chi^a$ variables; their $\star$-product is:
\be\label{starprod}
A\star B:=\exp\left[-\frac{\im}{2}\epsilon_{ijk}\,\gamma_{i}\,q^{j}[A]\,q^{k}[B]\right]\,A\,B\,,
\ee
with the product $AB$ on the right-hand side given by the usual Grassmannian multiplication. The three real parameters $\gamma_i$ appearing in the exponential phase are those of the $\gamma$-deformation.

This is applied to the twistor action of $\cN=4$ SYM by replacing the wedge product of differential forms on $\PT$ with the $\star$-product. In particular, we consider the action
\be\label{gTA}
S^{\gamma}[\cA]=S^{\gamma}_{1}[\cA]+\rg^2\,S^{\gamma}_{2}[\cA]\,,
\ee
where the two terms are given by\footnote{This version of $S^{\gamma}_1[\cA]$ first appeared in~\cite{Kulaxizi:2004pa} as the effective action of twistor-string theory deformed by the $\star$-product.}
\be\label{gTA1}
S^{\gamma}_{1}[\cA]=\frac{\im}{2\,\pi}\,\int_{\PT}\D^{3|4}Z\wedge \tr\left(\cA\star\dbar\cA+\frac{2}{3}\cA\star\cA\star\cA\right)\,,
\ee
and
\be\label{gTA2}
S^{\gamma}_{2}[\cA]=\oint_{\R^{4|8}}\d^{4|8}X\,\mathrm{log}\,\mathrm{det}_{\star}\!\left(\dbar+\cA\right)|_{X}\,.
\ee
In $S^{\gamma}_2$, the $\star$-deformed log det$(\dbar+\cA)|_X$ is understood through its perturbative expansion:
\begin{multline}\label{*logdet}
\mathrm{log}\,\mathrm{det}_{\star}\!\left(\dbar+\cA\right)|_{X}:=\tr\left(\log \dbar|_X\right) \\
+\sum_{n=2}^{\infty}\frac{1}{n}\left(\frac{1}{2\pi\im}\right)^{n} \int\limits_{(\P^1)^{n}} \frac{\D\sigma_{1}\cdots\D\sigma_{n}}{(1\,2)\,(2\,3)\cdots (n\,1)}\,\tr\left(\cA_1\star\cA_2\star\cdots\star\cA_n\right)\,.
\end{multline}
Crucially, the twistor superfield $\cA$ remains unchanged; its $\chi$-expansion is still given by \eqref{4TF}, containing only the usual Grassmann product. The $\gamma$-deformed twistor action remains invariant under non-abelian gauge transformations \eqref{4gi}.

\medskip

At this point, we have established that the twistorial $\gamma$-deformation defined by \eqref{starprod} leads to a deformed twistor action which is gauge-invariant. However, it is not clear that this $\gamma$-deformed twistor action is actually related to $\gamma$-deformed SYM on space-time. This is established by the following:

\begin{propn}\label{gTAprop}
The $\gamma$-deformed twistor action \eqref{gTA} is perturbatively equivalent to $\gamma$-deformed SYM theory on space-time, in the sense that solutions to its field equations are in one-to-one correspondence with solutions to the field equations of $\gamma$-deformed SYM, up to space-time gauge transformations. 
\end{propn}

\proof The proof is virtually equivalent to that for the undeformed $\cN=4$ SYM twistor action~\cite{Mason:2005zm,Boels:2006ir,Koster:2017fvf}. It suffices to show that with a partial gauge fixing -- which reduces the twistor space gauge freedom \eqref{4gi} to space-time gauge transformations -- the twistor action \eqref{gTA} is perturbatively equivalent to the space-time action of $\gamma$-deformed SYM. The relevant condition is the Woodhouse harmonic gauge~\cite{Woodhouse:1985id}:
\be\label{Wgauge}
\dbar^{*}|_{X}\cA|_X = 0\,,
\ee
where $\dbar^{*}|_{X}$ is the adjoint of the $\dbar$-operator restricted to any twistor line $X\cong\P^1$. Since $\dbar|_X \cA_X=0$ on dimensional grounds, \eqref{Wgauge} forces the twistor field $\cA$ to be harmonic upon restriction to the $\P^1$ fibres of twistor space. Residual gauge transformations which preserve \eqref{Wgauge} are harmonic functions on the fibres of twistor space, valued in the adjoint of the gauge group:
\begin{equation*}
\dbar^{*}|_{X}\dbar|_{X}g(Z) =0\,.
\end{equation*}
The only harmonic functions on $\P^1$ which are homogeneous of degree zero are constant, so the residual gauge freedom associated with \eqref{Wgauge} is precisely that of space-time gauge transformations: $g(Z)=g(x)$. 

We must now evaluate the $\gamma$-deformed twistor action \eqref{gTA} in the Woodhouse harmonic gauge. Expanding $\cA$ in the basis \eqref{eucb2} with the Woodhouse gauge condition, each component on the bosonic twistor space takes the form:
\be\label{Wgcomps}
a=a_{\dot\alpha}(x,\lambda,\hat{\lambda})\,\bar{e}^{\dot\alpha}\,, \qquad \tilde{\psi}_{a}=\tilde{\psi}_{a\dot\alpha}(x,\lambda,\hat{\lambda})\,\bar{e}^{\dot\alpha}\,,
\ee
\begin{equation*}
 \phi_{ab}=\Phi_{ab}(x)\,\bar{e}^{0}+\phi_{ab\dot\alpha}(x,\lambda,\hat{\lambda})\,\bar{e}^{\dot\alpha}\,, \qquad \psi^{a}=2\,\frac{\Psi^{a\alpha}(x)\,\hat{\lambda}_{\alpha}}{\la\lambda\,\hat{\lambda}\ra}\,\bar{e}^{0}+\psi^{a}_{\dot\alpha}(x,\lambda,\hat{\lambda})\,\bar{e}^{\dot\alpha}\,,
\end{equation*}
\begin{equation*}
 b=3\,\frac{B_{\alpha\beta}(x)\,\hat{\lambda}^{\alpha}\hat{\lambda}^{\beta}}{\la\lambda\,\hat{\lambda}\ra^2}\,\bar{e}^0+b_{\dot\alpha}(x,\lambda,\hat{\lambda})\,\bar{e}^{\dot\alpha}\,,
\end{equation*}
with all functional coefficients valued in the adjoint of the gauge group. 

First, consider $S^{\gamma}_{1}[\cA]$ in this gauge. Integration over the fermionic directions of twistor space annihilates all terms which are not proportional to $(\chi)^4$ after evaluating the $\star$-product. Using the definition \eqref{starprod}, it follows that the only contribution to $S^{\gamma}_{1}[\cA]$ for which the $\star$-product can lead to non-trivial phases is:
\be\label{Pproof1}
\DZ \text{tr}\left(\chi^{a}\Tilde{\psi}_{a}\star\chi^{b}\chi^{c}\phi_{bc}\star\chi^{d}\Tilde{\psi}_{d}\right)\,.
\ee
In the sum over R-symmetry indices, there are three distinct cases. The first is
\be\label{Pproof2}
\frac{1}{3}\DZ \text{tr}\left(\chi^{i}\Tilde{\psi}_{i}\star\chi^{j}\chi^{k}\phi_{jk}\star\chi^{4}\Tilde{\psi}_{4}\right)
= -\frac{\epsilon^{ijk}}{72}\int\D^{3}Z\wedge \text{tr}\left(\Tilde{\psi}_{i}\wedge\phi_{jk}\wedge\Tilde{\psi}_{4}\right) \e^{-\frac{\im}{2}\gamma_{i}^{-}}\,,
\ee
where the combination $\gamma_{i}^{\pm}$ is defined as
\be\label{gammapmdef}
\gamma_{1}^{\pm}:=-\frac{(\gamma_3\pm\gamma_2)}{2}\,, \qquad \gamma^{\pm}_2:=-\frac{(\gamma_1\pm\gamma_3)}{2}\,, \qquad \gamma_{3}^{\pm}:=-\frac{(\gamma_2\pm\gamma_1)}{2}\,.
\ee
Similarly, the second and third cases are
\be\label{Pproof3}
\frac{1}{3}\DZ \text{tr}\left(\chi^{4}\Tilde{\psi}_{4}\star\chi^{j}\chi^{k}\phi_{jk}\star\chi^{l}\Tilde{\psi}_{l}\right)
= -\frac{\epsilon^{jkl}}{72}\int\D^{3}Z\wedge \text{tr}\left(\Tilde{\psi}_{4}\wedge\phi_{jk}\wedge\Tilde{\psi}_{l}\right) \e^{\frac{\im}{2}\gamma_{l}^{-}}\,,
\ee
and
\be\label{Pproof4}
\begin{split}
&\frac{2}{3}\DZ \text{tr}\left(\chi^{i}\Tilde{\psi}_{i}\star\chi^{j}\chi^{4}\phi_{j4}\star\chi^{k}\Tilde{\psi}_{k}\right)\\
& = \frac{\epsilon^{ijk}}{36}\,\int\D^{3}Z\wedge \text{tr}\left(\Tilde{\psi}_{i}\wedge\phi_{j4}\wedge\Tilde{\psi}_{k}\right)\exp\left[\frac{\im}{2}\,\epsilon^{kim}\,\gamma_{m}^{+}\right]\,,
\end{split}
\ee
respectively. The three distinct cyclic permutations of the fields in the trace can be shown to contribute with the same phases.

Define the normalized (bosonic) holomorphic 3-form on twistor space
\be\label{omega}
\Omega:=\frac{\D^{3}Z}{4!}=\la\lambda\,\hat{\lambda}\ra^{4}\,e^{0}\wedge e^{\dot\alpha}\wedge e_{\dot\alpha}\,.
\ee
Combining \eqref{Pproof2} -- \eqref{Pproof4} with all the other terms in $S^{\gamma}_1[\cA]$ which are effectively undeformed by the $\star$-product, leaves:
\begin{multline}\label{Pproof5}
 \frac{\im}{2\,\pi}\int_{\PT}\frac{\Omega\wedge\Bar{\Omega}}{\la\lambda\,\hat{\lambda}\ra^4}\,\mathrm{tr}\left(b^{\dot\alpha}\,\dbar_0 a_{\dot\alpha}+3\frac{B_{\alpha\beta}\,\hat{\lambda}^{\alpha}\hat{\lambda}^{\beta}}{\la\lambda\,\hat{\lambda}\ra^2}\,\left(\dbar_{\dot\delta}a^{\dot\delta}-\frac{1}{2}[a_{\dot\delta},\,a^{\dot\delta}]\right) +\psi^{a\dot\alpha}\,\dbar_0\Tilde{\psi}_{a\dot\alpha}\right. \\
 +2\frac{\Psi^{a\alpha}\,\hat{\lambda}_{\alpha}}{\la\lambda\,\hat{\lambda}\ra}\left(\dbar_{\dot\delta}\Tilde{\psi}^{\dot\delta}_{a}+[a_{\dot\delta},\,\Tilde{\psi}^{\dot\delta}_{a}]\right)-\frac{\epsilon^{abcd}}{4}\,\Phi_{ab}\left(\dbar_{\dot\delta}\phi_{cd}^{\dot\delta}+[a_{\dot\delta},\,\phi_{cd}^{\dot\delta}]\right)+\frac{\epsilon^{ijk}}{2}\left(\Tilde{\psi}^{\dot\delta}_{i}\,\Phi_{jk}\,\Tilde{\psi}_{4\dot\delta}\,\e^{-\frac{\im}{2}\gamma_{i}^{-}}\right. \\
 \left.\left. -\Tilde{\psi}^{\dot\delta}_{4}\,\Phi_{jk}\,\Tilde{\psi}_{i\dot\delta}\,\e^{\frac{\im}{2}\gamma_{i}^{-}}\right)-\epsilon^{ijk}\,\Tilde{\psi}_{i}^{\dot\delta}\,\Phi_{j4}\,\Tilde{\psi}_{k\dot\delta}\,\exp\left[\frac{\im}{2}\,\epsilon^{kim}\,\gamma_{m}^{+}\right]-\frac{\epsilon^{abcd}}{2}\,\phi_{ab}^{\dot\alpha}\,\dbar_{0}\phi_{cd\dot\alpha}\right)\,.
\end{multline}
The functions $\psi^{a}_{\dot\alpha}$ and $b_{\dot\alpha}$ act as Lagrange multipliers, and $\phi_{ab}^{\dot\alpha}$ can also be integrated out. These impose that the remaining twistor functions in \eqref{Pproof5} obey
\begin{equation}\label{Pproof6}
a_{\Dot{\alpha}}(x,\lambda,\hat\lambda)=A_{\alpha\Dot{\alpha}}(x)\,\lambda^{\alpha}, \quad\Tilde{\psi}_{a\Dot{\alpha}}(x,\lambda,\hat\lambda)=\Tilde{\Psi}_{a\Dot{\alpha}}(x),\quad\phi_{ab\Dot\alpha}=\frac{\hat\lambda^{\alpha}}{\langle\lambda\,\hat\lambda\rangle}\,D_{\alpha\Dot{\alpha}}\Phi_{ab}(x)\,,
\end{equation}
where $D_{\alpha\dot\alpha}$ is the gauge covariant derivative on $\R^4$ defined by $A_{\alpha\dot\alpha}$. With these expressions, all dependence on the $\P^1$ fibre coordinates $(\lambda,\hat{\lambda})$ is manifest in \eqref{Pproof5}, and the integral over the fibre can be performed explicitly (cf., \cite{Boels:2006ir}). This leaves an action functional on $\R^4$:
\begin{equation}\label{Pproof7}
\begin{split}    
		S^{\gamma}_{1}[\cA]
		 &= \int\text{d}^{4}x\text{ tr}\left(-\frac12B_{\alpha\beta}\,F^{\alpha\beta}-\Psi^{a\alpha}\,D_{\alpha\Dot{\alpha}}\Tilde{\Psi}_{a}^{\Dot{\alpha}}+\frac{\epsilon^{abcd}}{8}\, D_{\alpha\Dot{\alpha}}\Phi_{ab}\,D^{\alpha\Dot{\alpha}}\Phi_{cd}\right. \\
		&\quad \left.  +\Tilde{\Psi}_{i}^{\Dot{\alpha}}\,\Phi^{i}\,\Tilde{\Psi}_{4\Dot{\alpha}}\,\e^{-\frac{\im}{2}\gamma^{-}_{i}}-\Tilde{\Psi}_{4}^{\Dot{\alpha}}\,\Phi^{i}\,\Tilde{\Psi}_{i\Dot{\alpha}}\,\e^{\frac{\im}{2}\gamma^{-}_{i}} -\epsilon^{ijk}\,\Tilde{\Psi}_{k}^{\Dot{\alpha}}\,\Phi_{i}^{\dagger}\,\Tilde{\Psi}_{j\Dot{\alpha}}\,\e^{\frac{\im}{2}\epsilon^{jkm}\gamma_{m}^{+}}\right)\,,
		\end{split}
\end{equation}
where $F_{\alpha\beta}$ is the anti-self-dual part of the field strength of $A_{\alpha\dot\alpha}$ and we have defined $\frac12\epsilon^{ijk}\Phi_{jk}\equiv\Phi^{i}$ and $\Phi_{i4}\equiv\Phi^{\dagger}_{i}$ for the space-time scalars.

\medskip

Now consider the non-local part of the twistor action, $S_{2}^{\gamma}[\cA]$. Woodhouse harmonic gauge leaves only the $n=2,3,4$ terms of the infinite sum \eqref{*logdet}, since the forms $a$ and $\tpsi_{a}$ have no component along the $\P^{1}$ fibres. In the integral over the space of lines in $\PT$, there is a SL$(2,\C)$ redundancy associated with the automorphism group of the line. We fix this redundancy by identifying the intrinsic homogeneous coordinates on each copy of $\P^1$ ($\sigma_i^{\alpha}$) fibre-wise with the projective coordinates $\lambda_i^{\alpha}$ on twistor space.  

With this identification, we can proceed to compute each of the $n=2,3,4$ contributions to $S^{\gamma}_{2}[\cA]$. The $n=2$ contribution is identical to the undeformed case~\cite{Boels:2006ir}:
\begin{equation}\label{*Pproof1}
\begin{split}
\frac{1}{2}\left(\frac{1}{2\pi \im}\right)^{2}\oint\text{d}^{4|8}X\int \frac{\text{D}\lambda_{1}\text{D}\lambda_{2}}{\langle12\rangle\langle21\rangle}\text{ tr}\left(\cA_{1}\star\cA_{2}\right)
& = -\frac{1}{2}\int\text{d}^{4}x\text{ tr}\left(B_{\alpha\beta}\,B^{\alpha\beta}\right)\,.
\end{split} 
\end{equation}
However, the $\star$-product acts non-trivially on the $n=3,4$ contributions. For the $n=3$ case, the only non-vanishing contributions arise from 
\begin{multline}\label{*Pproof2}
\text{tr}\left(\frac{\left(\chi_{1}\right)^{3}_{a}}{3!}\psi^{a}\star\frac{\chi^{b}_{2}\chi^{c}_{2}}{2}\phi_{bc}\star\frac{\left(\chi_{3}\right)^{3}_{d}}{3!}\psi^{d}\right) =\frac{1}{2}\left(\frac{1}{3!}\right)^{2}\text{tr}\left(\left(\chi_{1}\right)^{3}_{i}\,\psi^{i}\star\chi^{j}_{2}\chi^{k}_{2}\,\phi_{jk}\star\left(\chi_{3}\right)^{3}_{l}\,\psi^{l}\right. \\
\left.+2\,\left(\chi_{1}\right)^{3}_{i}\,\psi^{i}\star\chi^{j}_{2}\chi^{4}_{2}\,\phi_{j4}\star\left(\chi_{3}\right)^{3}_{4}\,\psi^{4} +2\,\left(\chi_{1}\right)^{3}_{4}\,\psi^{4}\star\chi^{j}_{2}\chi^{4}_{2}\,\phi_{j4}\star\left(\chi_{3}\right)^{3}_{l}\,\psi^{l}\right)\,.
\end{multline}
Evaluating the $\star$-product in each of these cases yields a phase, and once again all cyclic permutations of fields in the trace contribute with the same phase. Performing the $\P^1$ integrations using the methods of~\cite{Boels:2006ir} gives 
\begin{equation}\label{*Pproof3}
\int\text{d}^{4}x\text{ tr}\left(-\Psi^{i\alpha}\,\Phi_{i}^{\dagger}\,\Psi^{4}_{\alpha}\,\e^{-\frac{\im}{2}\gamma^{-}_{i}}+\Psi^{4\alpha}\,\Phi_{i}^{\dagger}\,\Psi^{i}_{\alpha}\,\e^{\frac{\im}{2}\gamma^{-}_{i}}+\epsilon_{ikl}\,\Psi^{i\alpha}\,\Phi^{k}\,\Psi^{l}_{\alpha}\,\e^{\frac{\im}{2}\epsilon_{lim}\gamma^{+}_{m}}\right)\,,
\end{equation}
for the $n=3$ contribution.

The $n=4$ contribution arises from four insertions of the twistor field $\phi$, which can occur in six distinct structures:
\begin{equation}\label{*Pproof4}
\begin{split}
&\frac{1}{16}\left(\frac{1}{2\pi i}\right)^{4}\oint\text{d}^{4|8}X\int \frac{\text{D}\lambda_{1}\text{D}\lambda_{2}\text{D}\lambda_{3}\text{D}\lambda_{4}}{\langle12\rangle\langle23\rangle\langle34\rangle\langle41\rangle}\text{tr} \left(\chi^{i}_{1}\chi^{j}_{1}\phi_{ij}\star\chi^{k}_{2}\chi^{l}_{2}\phi_{kl}\star\chi^{m}_{3}\chi^{4}_{3}\phi_{m4}\star\chi^{p}_{4}\chi^{4}_{4}\phi_{p4}\right. \\
&\phantom{}\left.+\chi^{i}_{1}\chi^{j}_{1}\phi_{ij}\star\chi^{k}_{2}\chi^{4}_{2}\phi_{k4}\star\chi^{l}_{3}\chi^{m}_{3}\phi_{lm}\star\chi^{p}_{4}\chi^{4}_{4}\phi_{p4}+\chi^{i}_{1}\chi^{4}_{1}\phi_{i4}\star\chi^{j}_{2}\chi^{k}_{2}\phi_{jk}\star\chi^{l}_{3}\chi^{m}_{3}\phi_{lm}\star\chi^{p}_{4}\chi^{4}_{4}\phi_{p4}\right. \\
&\left.+\chi^{i}_{1}\chi^{j}_{1}\phi_{ij}\star\chi^{k}_{2}\chi^{4}_{2}\phi_{k4}\star\chi^{l}_{3}\chi^{4}_{3}\phi_{l4}\star\chi^{m}_{4}\chi^{p}_{4}\phi_{mp}+\chi^{i}_{1}\chi^{4}_{1}\phi_{i4}\star\chi^{j}_{2}\chi^{k}_{2}\phi_{jk}\star\chi^{l}_{3}\chi^{4}_{3}\phi_{l4}\star\chi^{m}_{4}\chi^{p}_{4}\phi_{mp}\right. \\
&\left.+\chi^{i}_{1}\chi^{4}_{1}\phi_{i4}\star\chi^{j}_{2}\chi^{4}_{2}\phi_{j4}\star\chi^{k}_{3}\chi^{m}_{3}\phi_{km}\star\chi^{l}_{4}\chi^{p}_{4}\phi_{lp}\right)\,. 
\end{split}
\end{equation}
For each term, we evaluate the $\star$-product then use the result of~\cite{Koster:2017fvf} section 2.5 to perform the $\P^1$ integrals. The result for all of the $n=4$ contributions is remarkably simple:
\be\label{*Pproof5}
\int\text{d}^{4}x\text{ tr}\left(\Phi^{i}\Phi^{j}\Phi_{i}^{\dagger}\Phi_{j}^{\dagger}\exp\left[-\im\epsilon_{ijk}\gamma_{k}\right]-\frac{1}{4}\left\{\Phi_{i}^{\dagger},\,\Phi^{i}\right\}\left\{\Phi_{j}^{\dagger},\,\Phi^{j}\right\}\right)\,.
\ee
Combining \eqref{*Pproof1}, \eqref{*Pproof3} and \eqref{*Pproof4} gives:
\begin{multline}\label{*Pproof6}
S^{\gamma}_{2}[\cA]=\int\text{d}^{4}x\text{ tr}\left(-\frac{1}{2}B_{\alpha\beta}\,B^{\alpha\beta}-\Psi^{i\alpha}\,\Phi_{i}^{\dagger}\,\Psi^{4}_{\alpha}\,\e^{-\frac{\im}{2}\gamma^{-}_{i}}+\Psi^{4\alpha}\,\Phi_{i}^{\dagger}\,\Psi^{i}_{\alpha}\,\e^{\frac{\im}{2}\gamma^{-}_{i}}\right.\\
\left.+\epsilon_{ijk}\Psi^{i\alpha}\,\Phi^{j}\,\Psi^{k}_{\alpha}\,\e^{\frac{\im}{2}\epsilon_{kil}\gamma^{+}_{l}}+\Phi^{i}\,\Phi^{j}\,\Phi_{i}^{\dagger}\,\Phi_{j}^{\dagger}\,\e^{-\im\epsilon_{ijk}\gamma_{k}}-\frac{1}{4}\left\{\Phi_{i}^{\dagger},\,\Phi^{i}\right\}\,\left\{\Phi_{j}^{\dagger},\,\Phi^{j}\right\}\right)\,
\end{multline}
for $S_{2}^{\gamma}[\cA]$ evaluated in Woodhouse harmonic gauge.

The combination of $S_{1}^{\gamma}+\rg^2\,S_2^{\gamma}$ through \eqref{Pproof7}, \eqref{*Pproof6} defines an action on $\R^4$ with the standard non-abelian gauge invariance. Integrating out the field $B_{\alpha\beta}$ using its equations of motion, one obtains an action which differs from that of $\gamma$-deformed $\cN=4$ SYM by a multiple of the topological term $\int \tr(F\wedge F)$. Thus, the action is perturbatively equivalent to $\gamma$-deformed SYM. The simple field redefinition 
\be\label{Proofd}
\Phi^{i}\rightarrow\im\, \Phi^{i}\,, \quad \Psi^{i}\rightarrow\frac{\im}{\sqrt{\rg}}\,\Phi^{i}\,, \quad \tilde{\Psi}_{i}\rightarrow-\im\sqrt{\rg}\,\tilde{\Psi}_{i}\,, \quad \Phi^{\dagger}_{i}\rightarrow-\im\,\Phi^{\dagger}_{i}\,,
\ee
matches this action with a form that often appears in the literature (e.g., \cite{Fokken:2013aea,Gurdogan:2015csr}). \qed


\section{Double scaling limit in twistor space}
\label{DSLIM}

In~\cite{Gurdogan:2015csr,Caetano:2016ydc}, a double scaling limit of $\gamma$-deformed SYM was considered, where the deformation parameters $\gamma_i\rightarrow \im\infty$ for $i=1,2,3$, the gauge coupling constant $\rg\rightarrow 0$, and the three effective couplings $\xi_{i}:=\rg\e^{-\frac{\im}{2}\gamma_i}$ are left finite. The gauge field decouples in this limit, and the resulting theory is a non-unitary (since the deformation parameters -- and thus the action -- are complexified) but extremely simple theory of scalars and fermions with only cubic and quartic interactions. This theory is known as four-dimensional \emph{chiral field theory} ($\chi$FT), which is believed to be integrable at the quantum level in the planar limit~\cite{Caetano:2016ydc,Kazakov:2018hrh}. Conformal fishnet theory (FCFT) is obtained from $\chi$FT by setting to zero two of the effective couplings, leaving only a quartic theory of two complex scalars.

In this section, we implement the double scaling limit in twistor space, starting from the $\gamma$-deformed twistor action. The resulting twistor action is shown to give a classical (non-perturbative) description of $\chi$FT, and upon setting to zero two effective couplings, gives a classical (non-perturbative) description of FCFT.


\subsection{Double scaling limit and $\chi$FT}

On $\PT$, the double scaling limit can be implemented directly through the $\star$-product and gauge coupling. However, direct comparison with the space-time action requires rescaling some space-time fields by powers of $\sqrt{\rg}$, as in \eqref{Proofd}. This can be accomplished in twistor space by simultaneously rescaling the fermionic coordinates on $\PT$ as well as the twistor field $\cA$:
\be\label{ds1}
\chi^{a}\rightarrow \sqrt{\rg}\,\chi^{a}\,, \qquad \cA\rightarrow \rg\,\cA\,.
\ee
Compatibility between these two rescalings implies that the bosonic components of $\cA$ are rescaled as:
\be\label{ds2}
a\rightarrow \rg\,a\,, \quad \tilde{\psi}_{a}\rightarrow \sqrt{\rg}\,\tilde{\psi}_a\,, \quad \phi_{ab}\rightarrow \phi_{ab}\,, \quad \psi^{a}\rightarrow \rg^{-\frac{1}{2}}\,\psi^{a}\,, \quad b\rightarrow \rg^{-1}\,b\,.
\ee
The holomorphic measure on $\PT$ and the measure on the space of twistor lines transform by
\be\label{ds3}
\D^{3|4}Z\rightarrow \rg^{-2}\,\D^{3|4}Z\,, \qquad \d^{4|8}X\rightarrow \rg^{-4}\,\d^{4|8}X\,,
\ee
respectively under this rescaling.

As a result, the two terms in the $\gamma$-deformed twistor action are trivially rewritten as
\be\label{dsTA1}
S^{\gamma}_{1}[\cA]=\frac{\im}{2\,\pi}\,\int_{\PT}\D^{3|4}Z\wedge \tr\left(\cA\star\dbar\cA+\frac{2\,\rg}{3}\,\cA\star\cA\star\cA\right)\,,
\ee
and
\be\label{dsTA2}
S^{\gamma}_{2}[\cA]=\rg^{-4}\,\oint_{\R^{4|8}}\d^{4|8}X\,\mathrm{log}\,\mathrm{det}_{\star}\!\left(\dbar+\rg\,\cA\right)|_{X}\,.
\ee
In the double scaling limit 
\be\label{dslimit}
\rg\rightarrow 0\,, \qquad \gamma_{i}\rightarrow \im\,\infty\,, \qquad \xi_{i}:=\rg\,\e^{-\frac{\im}{2}\gamma_i}=\mbox{ finite}\,,
\ee
it is easy to see that all quadratic terms in $S^{\gamma}[\cA]$ remain finite (and undeformed), so we need only consider the interacting terms. It is useful to make the identification 
\be\label{scalarlabel}
\phi_{i}^{\dagger}:=\phi_{i4}\,, \qquad \phi^{i}:=\frac{1}{2}\epsilon^{ijk}\,\phi_{jk}\,,
\ee
for the weight $-2$ components of $\cA$.

Since the cubic interaction in $S^{\gamma}_{1}[\cA]$ scales as $\rg$, only terms which are proportional to $\e^{-\frac{\im}{2}\gamma}$ (for some combination $\gamma$ of the $\gamma_i$s) will survive in the double scaling limit. Using the definition of the $\star$-product, it is easy to identify such terms:
\begin{multline}\label{dsTA11}
\lim_{\substack{\rg\rightarrow0 \\ \gamma_i\rightarrow\im\infty}} \frac{\im\,\rg}{3\,\pi}\,\int_{\PT}\D^{3|4}Z\wedge\tr\left(\cA\star\cA\star\cA\right) =\frac{\im}{2\,\pi}\, \int_{\PT}\D^{3}Z\wedge \tr\left(\sqrt{\xi_1\xi_2}\,\tilde{\psi}_{2}\wedge \phi^{\dagger}_{3}\wedge\tilde{\psi}_{1} \right. \\
\left.+\sqrt{\xi_1\xi_3}\,\tilde{\psi}_1\wedge\phi^{\dagger}_{2}\wedge\tilde{\psi}_{3} +\sqrt{\xi_2\xi_3}\,\tilde{\psi}_3\wedge\phi^{\dagger}_{1}\wedge\tilde{\psi}_2\right)\,.
\end{multline}
In taking the limit, integration over the fermionic directions of twistor space has been performed explicitly; we abuse notation by denoting the bosonic twistor space (an open subset of $\P^3$ with homogeneous coordinates $Z^A$) as $\PT$.

To account for contributions from $\rg^2 S^{\gamma}_2[\cA]$, one expands $\mathrm{log}\,\mathrm{det}_{\star}\!\left(\dbar+\rg\,\cA\right)|_{X}$. Only terms proportional to $(\rg\e^{-\frac{\im}{2}\gamma})^{k}$ for some integer $k>2$ will survive in the double scaling limit. It is easy to see that there are non-vanishing cubic and quartic contributions of this type:
\begin{multline}\label{dsTA21}
 \left.\left(\lim_{\substack{\rg\rightarrow0 \\ \gamma_i\rightarrow\im\infty}}\rg^2\,S^{\gamma}_{2}[\cA]\right)\right|_{O(\cA^3)}= 
 \oint_{\R^4}\d^4 X\,\int_{(\P^1)^{3}} \frac{\D\sigma_1\,\D\sigma_2\,\D\sigma_3}{16\,(2\pi\im)^3}\,(1\,3) \\
 \mathrm{tr}\left(\sqrt{\xi_1\xi_2}\,\psi^{2}_1\,\phi^{3}_2\,\psi^{1}_3+\sqrt{\xi_1\xi_3}\,\psi^{1}_1\,\phi^{2}_2\,\psi^{3}_3 +\sqrt{\xi_2\xi_3}\,\psi^{3}_1\,\phi^{1}_2\,\psi^{2}_3\right)\,,
\end{multline}
and
\begin{multline}\label{dsTA22}
 \left.\left(\lim_{\substack{\rg\rightarrow0 \\ \gamma_i\rightarrow\im\infty}}\rg^2\,S^{\gamma}_{2}[\cA]\right)\right|_{O(\cA^4)}= 
 \oint_{\R^4}\d^4 X\,\int_{(\P^1)^{4}} \frac{\D\sigma_1\,\D\sigma_2\,\D\sigma_3\,\D\sigma_4}{(2\pi\im)^4}\,\mathrm{tr}\left(\xi_1^2\,(\phi^{\dagger}_2)_1\,(\phi^{\dagger}_3)_{2}\,\phi^{2}_3\,\phi^{3}_4\right. \\
 +\left.\xi_{2}^{2}\,(\phi^{\dagger}_3)_1\,(\phi^{\dagger}_1)_2\,\phi^{3}_3\,\phi^{1}_{4} +\xi^2_3\,(\phi^{\dagger}_1)_1\,(\phi^{\dagger}_2)_2\,\phi^{1}_3\,\phi^{2}_4\right)\,.
\end{multline}
Here, subscripts on twistor fields indicate the fibre dependence of that field: $\psi^{2}_1=\psi^{2}(Z(\sigma_1))$, and so forth. All higher-order terms in the expansion of $S^{\gamma}_2$ are easily seen to vanish in the double-scaling limit.

\medskip

Collecting these results, the twistor action for the double scaling limit of $\gamma$-deformed $\cN=4$ SYM is given by two pieces: 
\be\label{dsTA}
S^{\mathrm{DS}}[\tilde{\psi}_i,\phi^{i},\psi^i]=S_1^{\mathrm{DS}}+S_2^{\mathrm{DS}}\,,
\ee
where $S^{\mathrm{DS}}_1$ is local on $\PT$ and $S^{\mathrm{DS}}_2$ is non-local. In detail, these are given by
\begin{multline}\label{dsTA1*}
 S^{\mathrm{DS}}_1=\frac{\im}{2\,\pi}\int_{\PT}\D^{3}Z\wedge \tr\left(\phi^{\dagger}_i\wedge\dbar\phi^{i}+\tilde{\psi}_{i}\wedge\dbar\psi^{i} +\sqrt{\xi_1\xi_2}\,\tilde{\psi}_{2}\wedge \phi^{\dagger}_{3}\wedge\tilde{\psi}_{1} \right.  \\
\left.+\sqrt{\xi_1\xi_3}\,\tilde{\psi}_1\wedge\phi^{\dagger}_{2}\wedge\tilde{\psi}_{3} +\sqrt{\xi_2\xi_3}\,\tilde{\psi}_3\wedge\phi^{\dagger}_{1}\wedge\tilde{\psi}_2\right)\,,
\end{multline}
and
\begin{multline}\label{dsTA2*}
 S^{\mathrm{DS}}_2=\oint_{\R^4}\d^4 X\,\int_{(\P^1)^{3}} \frac{\D\sigma_1\,\D\sigma_2\,\D\sigma_3}{16\,(2\pi\im)^3}\,(1\,3)\,\mathrm{tr}\left(\sqrt{\xi_1\xi_2}\,\psi^{2}_1\,\phi^{3}_2\,\psi^{1}_3+\sqrt{\xi_1\xi_3}\,\psi^{1}_1\,\phi^{2}_2\,\psi^{3}_3\right. \\
 \left.+\sqrt{\xi_2\xi_3}\,\psi^{3}_1\,\phi^{1}_2\,\psi^{2}_3\right) +\oint_{\R^4}\d^4 X\,\int_{(\P^1)^{4}} \frac{\D\sigma_1\,\D\sigma_2\,\D\sigma_3\,\D\sigma_4}{(2\pi\im)^4}\,\mathrm{tr}\left(\xi_1^2\,(\phi^{\dagger}_2)_1\,(\phi^{\dagger}_3)_{2}\,\phi^{2}_3\,\phi^{3}_4\right. \\
 +\left.\xi_{2}^{2}\,(\phi^{\dagger}_3)_1\,(\phi^{\dagger}_1)_2\,\phi^{3}_3\,\phi^{1}_{4} +\xi^2_3\,(\phi^{\dagger}_1)_1\,(\phi^{\dagger}_2)_2\,\phi^{1}_3\,\phi^{2}_4\right)
\end{multline}
While the fields $a$ and $b$ -- which correspond to gauge degrees of freedom on space-time -- have decoupled, the twistor action \eqref{dsTA} retains a purely twistorial gauge invariance under
\be\label{dsGI}
\tilde{\psi}_i\rightarrow \tilde{\psi}_i+\dbar \tilde{\beta}_i\,, \qquad \phi^i\rightarrow \phi^i+\dbar\alpha^i\,, \qquad \psi^i\rightarrow\psi^i+\dbar\beta^i\,,
\ee
where $\{\tilde{\beta}_i,\alpha^i,\beta^i\}$  are adjoint-valued functions on $\PT$ of weight $-1$, $-2$ and $-3$, respectively.

This twistor action describes a space-time field theory of three complex scalars and fermions, known as the \emph{chiral field theory} ($\chi$FT)~\cite{Gurdogan:2015csr,Caetano:2016ydc}. Unlike the $\gamma$-deformation, the twistorial description of $\chi$FT is classically exact (i.e., non-perturbative), as the gauge field degrees of freedom are decoupled:

\begin{propn}\label{dsTAprop}
The double-scaling limit twistor action \eqref{dsTA} is equivalent to $\chi$FT, in the sense that solutions to its field equations are in one-to-one correspondence with solutions to the field equations of $\chi$FT. Furthermore, the twistor action and $\chi$FT actions take the same values when evaluated on corresponding field configurations. 
\end{propn}

\proof Using the gauge freedom \eqref{dsGI}, the twistor fields can be put into Woodhouse harmonic gauge
\be\label{dsprop1}
\dbar^{*}|_{X}\tilde{\psi}_i|_{X}=0\,, \qquad \dbar^{*}|_{X}\phi^i|_{X}=0\,, \qquad \dbar^{*}|_{X}\psi^i|_{X}=0\,.
\ee
The remaining gauge freedom on $\PT$ is then reduced to transformations \eqref{dsGI} for which $\{\tilde{\beta}_i,\alpha^i,\beta^i\}$ are harmonic functions when restricted to $X\cong\P^1$. But there are no such functions, since each of $\{\tilde{\beta}_i,\alpha^i,\beta^i\}$ have negative homogeneity on $\P^1$. Therefore, the Woodhouse gauge \eqref{dsprop1} leaves no residual gauge freedom on space-time.

Using the Woodhouse gauge condition,
\be\label{dsWgcomps}
\tilde{\psi}_{i}=\tilde{\psi}_{i\dot\alpha}(x,\lambda,\hat{\lambda})\,\bar{e}^{\dot\alpha}\,, \qquad \phi^i=\Phi^i(x)\,\bar{e}^{0}+\phi^{i}_{\dot\alpha}(x,\lambda,\hat{\lambda})\,\bar{e}^{\dot\alpha}\,,
\ee
\begin{equation*}
\psi^{i}=2\,\frac{\Psi^{i\alpha}(x)\,\hat{\lambda}_{\alpha}}{\la\lambda\,\hat{\lambda}\ra}\,\bar{e}^{0}+\psi^{i}_{\dot\alpha}(x,\lambda,\hat{\lambda})\,\bar{e}^{\dot\alpha}\,,
\end{equation*}
the proof now proceeds in the same way as Proposition \ref{gTAprop}. After integrating out all of the $\P^1$-fibre dependence from the twistor action, one finds:
\begin{multline}\label{dsprop2}
 S^{\mathrm{DS}}_1=\int_{\R^4}\d^4x\,\tr\left(\frac{1}{2}\partial_{\alpha\dot\alpha}\Phi^{\dagger}_{i}\,\partial^{\alpha\dot\alpha}\Phi^{i}-\Psi^{i\alpha}\,\partial_{\alpha\dot\alpha}\tilde{\Psi}_{i\dot\alpha} +\sqrt{\xi_1\xi_2}\,\tilde{\Psi}^{\dot\alpha}_2 \Phi^{\dagger}_3 \tilde{\Psi}_{1\dot\alpha} \right. \\
 +\sqrt{\xi_1\xi_3}\,\tilde{\Psi}^{\dot\alpha}_1 \Phi^{\dagger}_2 \tilde{\Psi}_{3\dot\alpha} +\sqrt{\xi_2\xi_3}\,\tilde{\Psi}^{\dot\alpha}_3 \Phi^{\dagger}_1 \tilde{\Psi}_{2\dot\alpha}\bigg)\,,
\end{multline}
\begin{multline}\label{dsprop3}
 S^{\mathrm{DS}}_2=\int_{\R^4}\d^4x\,\tr\left(\xi_1^2\,\Phi_2^{\dagger}\Phi_{3}^{\dagger}\Phi^{2}\Phi^{3}+\xi_2^2\,\Phi_3^{\dagger}\Phi_{1}^{\dagger}\Phi^{3}\Phi^{1} + \xi_3^2\,\Phi_1^{\dagger}\Phi_{2}^{\dagger}\Phi^{1}\Phi^{2}\right. \\
 \left. -\sqrt{\xi_1\xi_2}\,\Psi^{2\alpha} \Phi^3 \Psi^1_{\alpha}-\sqrt{\xi_1\xi_3}\,\Psi^{1\alpha} \Phi^2 \Psi^3_{\alpha}-\sqrt{\xi_2\xi_3}\,\Psi^{3\alpha} \Phi^1 \Psi^{2}_{\alpha}\right)\,.
\end{multline}
This agrees with the space-time action of $\chi$FT~\cite{Gurdogan:2015csr,Caetano:2016ydc} after performing the rescalings $\Psi^i\rightarrow\im\Psi^i$, $\Phi^{i}\rightarrow\im\Phi^i$, $\tilde{\Psi}_{i}\rightarrow -\im\tilde{\Psi}_i$ and $\Phi^{\dagger}_{i}\rightarrow-\im\Phi^{\dagger}_i$. \qed


\subsection{Classical conformal fishnet theory}

On space-time, FCFT is obtained directly from $\chi$FT by setting to zero two of the effective couplings, say $\xi_{1},\xi_{2}\rightarrow 0$. This decouples all of the fermions from $\chi$FT as well as one of the scalars, leaving a theory with only a single quartic interaction: 
\be\label{stCFT}
S[\varphi_1, \varphi_2]=\int \d^{4}x\:\tr\!\left(\partial^{\mu}\bar{\Phi}_1\,\partial_{\mu}\Phi^{1}+\partial^{\mu}\bar{\Phi}_2\,\partial_{\mu}\Phi^{2}+\xi^2\,\bar{\Phi}_1\bar{\Phi}_2\Phi^1\Phi^2\right)\,,
\ee
where $\xi_{3}:=\xi$.

Setting $\xi_{1},\xi_2=0$ in the twistor action for $\chi$FT \eqref{dsTA} leaves a remarkably simple theory on twistor space, with local kinetic terms and a non-local quartic interaction:
\begin{multline}\label{cfTA}
S[\phi_1,\phi_2]=\frac{\im}{2\,\pi}\int_{\PT}\D^{3}Z\wedge\tr\left(\phi^{\dagger}_1\wedge\dbar\phi^{1}+\phi^{\dagger}_2\wedge\dbar\phi^{2}\right) \\
+\xi^2\,\oint_{\R^4}\d^{4}X\,\int_{(\P^1)^4}\frac{\D\sigma_1\,\D\sigma_2\,\D\sigma_3\,\D\sigma_4}{(2\pi\im)^4}\,\mathrm{tr}\left((\phi^{\dagger}_1)_1\,(\phi^{\dagger}_2)_2\,\phi^{1}_3\,\phi^{2}_4\right)\,.
\end{multline}
It is easy to see that this action is invariant under global SU$(N)$ and U$(1)\times$U$(1)$ transformations of the twistor fields, but in addition the action is preserved by the local twistor gauge transformations:
\be\label{cfgauge}
\phi^{1,2}\rightarrow \phi^{1,2}+\dbar\alpha^{1,2}\,, \qquad \alpha^{1,2}\in\Omega^{0}(\PT,\cO(-2)\otimes\mathfrak{g})\,.
\ee
This freedom can be used to put the twistor fields in Woodhouse harmonic gauge
\be\label{cfWood}
\dbar^{*}|_{X}\phi^{1,2}|_{X}=0\,,
\ee
whence the twistor action \eqref{cfTA} reduces to the space-time action \eqref{stCFT} for FCFT.

\medskip

This establishes that classical FCFT can be obtained exactly (i.e., non-perturbatively) by lifting the $\gamma$-deformation and double scaling limit directly to twistor space. Although the double-scaling limit decouples all local gauge freedom on space-time, in twistor space there is always a local `abelian' SU$(N)$ symmetry. While crucial for establishing the equivalence between the twistor actions and space-time theories for both $\chi$FT and FCFT, the local gauge freedom on twistor space enables other gauge-fixings particularly amenable to performing calculations.


\section{Conformal fishnet theory in twistor space}
\label{FRULES}

At this point, we turn our focus to the study of perturbative conformal fishnet theory using the twistor action \eqref{cfTA}, particularly in the planar limit $N\rightarrow\infty$ of the SU$(N)$-valued scalar fields. As our subsequent focus will be the computation of scattering amplitudes in FCFT, we first recall the structure of these amplitudes in twistor space before describing the twistor Feynman rules of the theory and discussing the structure of UV-divergences and their removal by double trace counter-terms in twistor space.  


\subsection{Cohomological amplitudes}

The external legs of any scattering process in FCFT are given by on-shell, massless SU$(N)$-valued scalar fields. On space-time, these external legs are often represented with a momentum eigenstate basis, $\e^{\im k\cdot x}$ where $k^2=0$, but of course any basis of on-shell solutions will do. On twistor space, the Penrose transform \eqref{PenTran} means that any such on-shell external scalar field is represented by a cohomology class:
\be\label{ca1}
\Phi_{\mathrm{on-shell}}(x)\leftrightarrow\phi\in H^{0,1}(\PT,\cO(-2)\otimes\mathfrak{g})\,.
\ee
This cohomology class encodes the quantum numbers of external states in the scattering process. 

For instance, in a momentum basis the on-shell $k_\mu$ is encoded by the spinors $k_{\alpha}\tilde{k}_{\dot\alpha}$, and the space-time external field
\begin{equation*}
 \Phi(x)=\sT^{\sa}\,\e^{\im k\cdot x}\,,
\end{equation*}
is represented by a twistor cohomology class
\be\label{ca2}
\phi(Z;k)=\sT^{\sa}\,\int \d s\,s\,\bar{\delta}^{2}\!\left(k-s\,\lambda\right)\,\e^{\im s\,[\mu\,\tilde{k}]}\,,
\ee
where $\sT^{\sa}$ is a generator of SU$(N)$. In \eqref{ca2}, the holomorphic delta function 
\be\label{delta2}
\bar{\delta}^2\!(\lambda):=\bigwedge_{\alpha=0,1} \d\overline{\lambda_{\beta}}\frac{\partial}{\partial\overline{\lambda_\beta}}\left(\frac{1}{\lambda_{\alpha}}\right)=\bigwedge_{\alpha=0,1} \dbar\left(\frac{1}{\lambda_{\alpha}}\right)\,,
\ee
is a $(0,2)$-distribution enforcing the vanishing of both components of its argument. One of the distributional form degrees is integrated against the scale parameter $s$, so that $\phi(Z;k)$ is a SU$(N)$-valued $(0,1)$-form on $\PT$ of weight $-2$, as required by the Penrose transform.

Thus, a scattering amplitude in FCFT can be represented as a functional of these twistor cohomology classes. The twistor version of the LSZ truncation procedure would then be given by a pairing between some integral kernel and these cohomology classes, with the integral kernel taking values in $\oplus_{i=1}^{n}H^{0,1}(\PT_i,\cO(-2)\otimes\mathfrak{g})^\vee$, where the duality is defined by the Hilbert space structure of $H^{0,1}(\PT,\cO(-2))$. This pairing removes the twistor dependence, leaving only a function of the relevant quantum numbers, as expected for a scattering amplitude.

Unfortunately, this pairing is non-local on twistor space and relies on an explicit choice of space-time signature~\cite{Eastwood:1981b}. A more useful pairing for our purposes is~\cite{Adamo:2011cb,Adamo:2013cra}
\be\label{capair}
H^{0,1}(\PT,\cO(-2)\otimes\mathfrak{g})\times H^{0,2}_{c}(\PT,\cO(-2)\otimes\mathfrak{g})\rightarrow \C\,, \qquad (\phi,\rho)\mapsto\int_{\PT}\D^3Z\wedge\,\tr\!\left(\phi\wedge\rho\right)\,,
\ee
where the subscript on $H^{0,2}_c$ denotes compact support. With this pairing, a $n$-point scattering amplitude on $\PT$ is represented by an integral kernel
\be\label{cakern}
A_n(Z_1,\ldots Z_n)\in\bigoplus_{i=1}^{n}H^{0,2}_c(\PT_i,\cO(-2)\otimes\mathfrak{g})\,,
\ee
where each $\PT_i$ is charted with homogeneous coordinates $Z_i$. The numerical (physical) scattering amplitude is uniquely obtained from this \emph{cohomological amplitude} by the pairing \eqref{capair} with external wavefunctions. Knowing the cohomological amplitude on twistor space is equivalent to knowing the physical amplitude in terms of external momenta. 


\subsection{Feynman rules: vertices and propagator}

\begin{figure}[t]
\centering
\includegraphics[scale=.4]{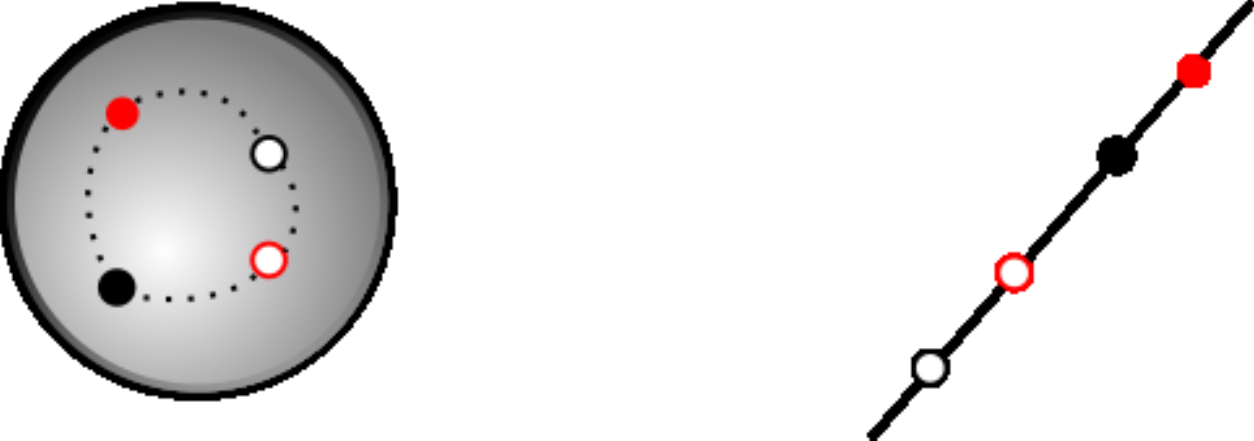}
\caption{The 4-point vertex in twistor space: insertions of $\phi^1$ and $\phi^2$ are denoted with filled black and red dots, insertions of $\phi^{\dagger}_1$ and $\phi^{\dagger}_2$ with black and red circles, respectively. The trace structure on the Riemann sphere (left) can be represented by arranging the insertions on a line (right).}
\label{Tvertex}
\end{figure}

The only vertex in the twistor action for classical conformal fishnet theory is given by the quartic interaction:
\be\label{vert1}
\xi^2\,\oint_{\R^4}\d^{4}X\,\int_{(\P^1)^4}\D\sigma_1\,\D\sigma_2\,\D\sigma_3\,\D\sigma_4\,\mathrm{tr}\left((\phi^{\dagger}_1)_1\,(\phi^{\dagger}_2)_2\,\phi^{1}_3\,\phi^{2}_4\right)
\ee
where the contour integral is over the moduli space of lines $X\cong\P^1$ in $\PT$ which are preserved by the Euclidean reality conditions\footnote{From now on, we will neglect factors of $(2\pi\im)$, as they may be viewed as implicit in the definitions of the projective integrals and distributions encountered.}. A pictorial version of this vertex is shown in figure~\ref{Tvertex}, with different operator insertions distinguished by their colour.

Now, a holomorphic linear map $Z^A:\P^1\rightarrow\PT$ can be written
\be\label{linmap1}
Z^A(\sigma)=X^{A}_{\alpha}\,\sigma^{\alpha}=A^{A}\,\sigma^{0}+B^A\,\sigma^1\,,
\ee
for $\sigma^{\alpha}$ the homogeneous coordinate on $\P^1$ and moduli $X^{A}_{\alpha}=\{A^A,\,B^A\}$. \emph{A priori}, there seem to be eight moduli, but the description of the linear map \eqref{linmap1} is redundant: we must account for the SL$(2,\C)$-invariance of $\P^1$ as well as the $\C^*$ projective rescalings of the target space.

Thus, the measure appearing in \eqref{vert1} can be written explicitly as
\be\label{vert2}
\d^4 X=\frac{\d^4 A\wedge\d^4 B}{\mathrm{vol}\;\GL(2,\C)}\,,
\ee
with the quotient by the (infinite) volume of GL$(2,\C)\cong\SL(2,\C)\times\C^*$ understood in the Fadeev-Popov sense. Writing the measure in this way has two benefits: first, it manifests the nature of the line $X\cong\P^1$ as the skew of two points $A,B\in\PT$. Secondly, this expression for the measure is manifestly conformally invariant, since it is constructed entirely from SL$(4,\C)$-invariant objects.

To construct cohomological amplitudes using this interaction, we must choose representatives for the four legs of this vertex. The suitable choice is given by \emph{elementary} states, which localize a field insertion on $X\cong\P^1$ to a point in twistor space:
\be\label{states}
\phi^{1,2}=\sT^{\sa}\,\bar{\delta}^{3}_{-2,-2}(Z,\,Z(\sigma))=\sT^{\sa} \int \d s\,s\,\bar{\delta}^{4}(Z+s\,Z(\sigma))\,,
\ee
where
\be\label{delta4}
\bar{\delta}^4(Z):=\bigwedge_{A=0}^{3} \dbar\left(\frac{1}{Z^A}\right)\,.
\ee
Thus, $\bar{\delta}^{3}_{-2,-2}(Z,Z(\sigma))$ is a $(0,3)$-distribution on $\PT$ enforcing the projective coincidence of its arguments, homogeneous of weight $-2$ in both $Z$ and $Z(\sigma)$. We often drop homogeneity subscripts on these distributional forms when their weights are clear from the context.

With the elementary states \eqref{states} and manifestly conformally invariant measure \eqref{vert2}, the four-point interaction vertex is $\xi^2\,\tr(\sT^{\sa_1}\cdots\sT^{\sa_4})\,V_4$, for
\be\label{vert3}
V_4(Z_1,Z_2,Z_3,Z_4):=\oint_{\R^4}\frac{\d^4 A\wedge\d^4 B}{\mathrm{vol}\;\GL(2,\C)}\,\int_{(\P^1)^4}\prod_{i=1}^{4}\D\sigma_i\,\bar{\delta}^{3}(Z_i,\,Z(\sigma_i))\,.
\ee
Integration over each of the four copies of $\P^1$ ensures that $V_4$ is a distributional $(0,2)$-form in each of the $\{Z_1,\ldots,Z_4\}$. The fact that $\{Z_i\}$ enter only through the projective delta functions makes it clear that $V_4$ is compactly supported. Finally, $\dbar V_4=0$ since
\be\label{vert4}
\dbar\,\bar{\delta}^{3}(Z_i,\,Z(\sigma_i))=0\,,
\ee
and the integrand of \eqref{vert3} is otherwise holomorphic. This establishes that
\be\label{vert5}
V_4(Z_1,\ldots,Z_4)\in\bigoplus_{i=1}^{4}H^{0,2}_c(\PT_i,\cO(-2))\,,
\ee
as required for the vertex in a cohomological representation. 

This statement is exact, in contrast to the situation for the vertices of the $\cN=4$ SYM twistor action, which fail to be $\dbar$-closed due to IR collinear divergences~\cite{Adamo:2011cb,Adamo:2013cra}. However, planar FCFT is free from IR-divergences, so \eqref{vert5} is a first reflection of this fact on twistor space. Finally, note that upon pairing the vertex using \eqref{capair} with momentum eigenstate representatives \eqref{ca2}, one obtains
\be\label{momrep}
\int_{\PT^4}V_4\,\prod_{i=1}^{4}\D^{3}Z_i\,\phi(Z_i;k_i)=\delta^{4}\!\left(\sum_{i=1}^{4}k_i\right)\,,
\ee
which is the standard momentum space, LSZ-truncated expression for the vertex.

In practical computations, the following facts about the 4-point vertex are extremely useful.
\begin{lemma}
 The cohomological 4-point vertex of FCFT obeys
 \be\label{vdecomp}
 \begin{split}
 V_{4}(1,2,3,4) & =\cV_3(1|2,3)\,\bar{\delta}^{2}_{-1,-1,-2}(2,3,4) \\
  & =\cV_2(2,3)\,\bar{\delta}^{2}_{-1,-1,-2}(2,3,1)\,\bar{\delta}^{2}_{-1,-1,-2}(2,3,4)\,,
 \end{split} 
 \ee
 where the pseudo-vertex $\cV_3$ is
 \begin{multline}\label{3vert}
 \cV_3(1|2,3):=\oint_{\R^4}\frac{\d^4 A\wedge\d^4 B}{\mathrm{vol}\;\GL(2,\C)}\,\int_{(\P^1)^3}(2\,3)\,\D\sigma_1\,\D\sigma_2\,\D\sigma_3\,\bar{\delta}^{3}_{-2,-2}(Z_1,\,Z(\sigma_1)) \\
 \bar{\delta}^{3}_{-1,-3}(Z_2,\,Z(\sigma_2))\,\bar{\delta}^{3}_{-1,-3}(Z_3,\,Z(\sigma_3))\,,
 \end{multline}
 taking values in
 \be\label{3vc}
 \cV_{3}(1|2,3)\in H^{0,2}_c(\PT_1,\cO(-2))\bigoplus_{i=2,3} H^{0,2}_c(\PT_i,\cO(-1))\,,
 \ee
 as a cohomological object. The pseudo-vertex $\cV_2$ is
 \be\label{2vert}
 \cV_2(2,3):=\oint\limits_{\R^4\times (\P^1)^2} \D^3A\wedge\D^3B\,\bar{\delta}^{3}_{0,-4}(Z_2,\,A)\,\bar{\delta}^{3}_{0,-4}(Z_3,\,B)\,,
 \ee
 where the contour integrates $A,B$ over the line $X\cong\P^1$ and then integrating the line over the real contour $\R^4\subset\M$, and $\cV_2$ takes values in
 \be\label{2vc}
 \cV_2(2,3)\in H^{0,2}_c(\PT_2,\cO)\oplus H^{0,2}_c(\PT_3,\cO)\,,
 \ee
 as a cohomological object.
\end{lemma}

\proof These relations follow from manipulation of the projective integrals over $\P^1$ in $V_4$. Let us work with affine coordinates on $\P^1$ (e.g., by choosing the coordinate patch where $\sigma_i^0\neq0$); we abuse notation, denoting these affine coordinates by $\sigma_i$ for $i=1,\ldots,4$. Writing out all scale integrals, the 4-point vertex is:
\be\label{pl1}
\oint_{\R^4}\frac{\d^4 A\wedge\d^4 B}{\mathrm{vol}\;\GL(2,\C)}\,\int_{(\C^*)^4}\prod_{i=1}^{4}\d\sigma_i\,\d s_{i}\,s_i\,\bar{\delta}^{4}(Z_i+s_i\,Z(\sigma_i))\,.
\ee
Change variables from $\sigma_4$ to $u$ via:
\be\label{pl2}
\sigma_4=\frac{s_2\,\sigma_2+u\,s_3\,\sigma_3}{s_2+u\,s_3}\,,
\ee
and observe that (in the affine coordinates)
\be\label{pl3}
Z(\sigma_i)=A+\sigma_i\,B\,,
\ee
whence \eqref{pl1} becomes
\begin{multline}
 \oint_{\R^4}\frac{\d^4 A\wedge\d^4 B}{\mathrm{vol}\;\GL(2,\C)}\,\int_{(\C^*)^4}\frac{\d u\,(\sigma_3-\sigma_2)}{(s_2+us_3)^2}\,\d s_4\,s_4\,\bar{\delta}^{4}\left(Z_4-\frac{s_4}{s_2+us_3}(Z_2+uZ_3)\right) \\
  \d\sigma_1\,\d s_1\,s_1 \bar{\delta}^{4}(Z_1+s_1 Z(\sigma_1))\,\prod_{i=2,3}\d\sigma_i\,\d s_i\,s_i^2\,\bar{\delta}^4(Z_i+s_i Z(\sigma_i))\,,
\end{multline}
making use of the various delta functions in play. Re-scaling $s_4\rightarrow (s_2+us_3)\,s_4$ and $Z_4\rightarrow s_4$ and restoring homogeneous coordinates on $\P^1$ leaves
\be\label{pl4}
\cV_3(1|2,3)\,\int_{(\C^*)^2}\d u\,\d s_4\,s_4\,\bar{\delta}^{4}(Z_2+uZ_3+tZ_4)=\cV_3(1|2,3)\,\bar{\delta}^{2}_{-1,-1,-2}(2,3,4)\,,
\ee
as claimed.

To obtain the second line of \eqref{vdecomp}, apply the same change of variables \eqref{pl2} now to $\sigma_1$ in $\cV_{3}(1|2,3)$. This results in
\begin{multline}\label{pl5}
\cV_{3}(1|2,3)=\oint_{\R^4}\frac{\d^4 A\wedge\d^4 B}{\mathrm{vol}\;\GL(2,\C)}\,\int_{(\P^1)^2} (2\,3)^2\, \D\sigma_2\,\D\sigma_3\,\bar{\delta}^{3}_{0,-4}(Z_2,\,Z(\sigma_2))\,\bar{\delta}^{3}_{0,-4}(Z_3,\,Z(\sigma_3)) \\
\times \bar{\delta}^{2}_{-1,-1,-2}(2,3,1)\,.
\end{multline}
The GL$(2,\C)$ redundancy in the measure can now be fixed by making the integrals over $A$ and $B$ projective (i.e., contracting with the Euler vector in $A$ and $B$) and setting $\sigma_{2}^{\alpha}=(1,0)$ and $\sigma_3^\alpha=(0,1)$, while removing the appropriate Jacobian factor. The result is the claimed expression \eqref{2vert}. \qed

\medskip

The kinetic operator for both $\phi^1$ and $\phi^2$ is the $\dbar$-operator on $\PT$; a useful form of its inverse has long been known in the context of the twistor action for $\cN=4$ SYM by using an axial gauge~\cite{Adamo:2011cb}. This axial gauge eliminates all vertices from the action except those arising from the non-local logdet$(\dbar+\cA)|_X$ term, which becomes a generating function for all MHV vertices of $\cN=4$ SYM. Although there is only one `MHV amplitude' in FCFT (the 4-point amplitude), we are still free to choose an axial gauge thanks to the twistor gauge invariance \eqref{cfgauge}.

In particular, the axial gauge is defined by a choice of fixed reference twistor $Z_*\in\PT$ such that the propagator $\Delta(Z_1,\,Z_2)$ obeys
\be\label{prop1}
\dbar\Delta(Z_1,\,Z_2)=\bar{\delta}^3_{-2,-2}(Z_1,\,Z_2)\,, \qquad Z_*\cdot\overline{\frac{\partial}{\partial Z_1}}\lrcorner \Delta=0=Z_*\cdot\overline{\frac{\partial}{\partial Z_2}}\lrcorner \Delta\,,
\ee
with $\Delta(Z_1,Z_2)$ understood to be a distribution with $(0,1)$-form degree in each of $Z_1$ and $Z_2$. This axial gauge can be imposed at the level of the twistor action through the usual gauge-fixing procedure, but (as usual for axial gauges) the ghost sector decouples from the path integral~\cite{Boels:2007qn}.

Building on the results for $\cN=4$ SYM, the propagator for each of the fields in FCFT on twistor space is:
\be\label{prop2}
\Delta(Z_1,\,Z_2)=\bar{\delta}^2_{-2,0,-2}(Z_1,\,*,\,Z_2)=\int \frac{\d s}{s}\,\d t\,t\,\bar{\delta}^4(Z_1+s\,Z_*+t\,Z_2)\,.
\ee
This is a $(0,2)$-distribution on $\PT_1\times\PT_2$ enforcing the projective collinearity of its three arguments. It is straightforward to show that \eqref{prop2} obeys the conditions of \eqref{prop1}; indeed, the calculation is just a simpler version of that for $\cN=4$ SYM. Firstly, one observes that
\be\label{prop3}
\dbar \Delta(Z_1,\,Z_2)=\int \d s\,\dbar\left(\frac{1}{s}\right)\,\d t\,t\,\bar{\delta}^4(Z_1+s\,Z_*+t\,Z_2)=\int \d t\,t\,\bar{\delta}^4(Z_1+t\,Z_2)=\bar{\delta}^3_{-2,-2}(Z_1,\,Z_2)\,,
\ee
so $\Delta$ is a Green's function for the $\dbar$-operator on twistor space. To establish that $\Delta$ is in the axial gauge defined by $Z_*$, one observes that
\begin{multline}\label{prop4}
 \Delta(Z_1,\,Z_2)=\int \frac{\d s}{s}\,\d t\,(1\hat{1}*\hat{2})\,(\hat{1}*2\hat{2})^2\,\dbar_1\left(\frac{1}{(1*2\hat{2})}\right)\,\dbar_2\left(\frac{1}{(2*1\hat{1})}\right) \\
 \times\bar{\delta}\left((1\hat{1}2\hat{2})+s(*\hat{1}2\hat{2})\right)\,\bar{\delta}\left((1\hat{1}2\hat{2})+t(2\hat{1}*\hat{2})\right) \\
 =-\frac{(1\hat{1}*\hat{2})\,(\hat{1}*2\hat{2})}{(1\hat{1}2\hat{2})}\,\dbar_1\left(\frac{1}{(1*2\hat{2})}\right)\,\dbar_2\left(\frac{1}{(2*1\hat{1})}\right)\,,
\end{multline}
where we employ the notation
\be\label{epsnot}
(1234):=\epsilon_{ABCD}\,Z_1^A\,Z_2^B\,Z_3^C\,Z_4^D\,,
\ee
and $\hat{Z}_1$, $\hat{Z}_2$ are the conjugates of $Z_1,\,Z_2$ with respect to the Euclidean reality condition. From the last line in \eqref{prop4}, it follows that
\begin{equation*}
 Z_*\cdot\overline{\frac{\partial}{\partial Z_1}}\lrcorner \Delta=0=Z_*\cdot\overline{\frac{\partial}{\partial Z_2}}\lrcorner \Delta\,,
\end{equation*}
due to the skew-symmetry of the 4-dimensional Levi-Civita symbol.

Finally, the colour structure of the propagator is the same as on space-time. Therefore, the full twistor space propagator for FCFT in axial gauge is:
\be\label{fullprop}
\Delta(Z_1,\,Z_2)=\bar{\delta}^2_{-2,0,-2}(Z_1,\,*,\,Z_2)\,\left(\delta^{\bar{j}_1}_{i_2}\,\delta^{\bar{j}_2}_{i_1}-\frac{1}{N}\,\delta^{\bar{j}_1}_{i_1}\,\delta^{\bar{j}_2}_{i_2}\right)\,,
\ee
where $i_1,i_2$ are fundamental indices of SU$(N)$ and $\bar{j}_1,\bar{j}_2$ are anti-fundamental indices associated with each end of the propagator.


\subsection{Divergences, counterterms and conformality}

It is well-known that the single-trace Lagrangian of FCFT is not quantum-complete due to UV divergences -- even in the planar limit~\cite{Fokken:2013aea,Sieg:2016vap}. These divergences come with a double-trace structure, and their removal necessitates adding double-trace counterterms to the space-time action. Of course, we could translate these counterterms to the twistor action, but it is more enlightening to derive them directly in twistor space. With the single-trace vertex \eqref{vert3} and propagator \eqref{fullprop} in twistor space, we have all the required tools.

\begin{figure}[t]
\centering
\includegraphics[scale=.7]{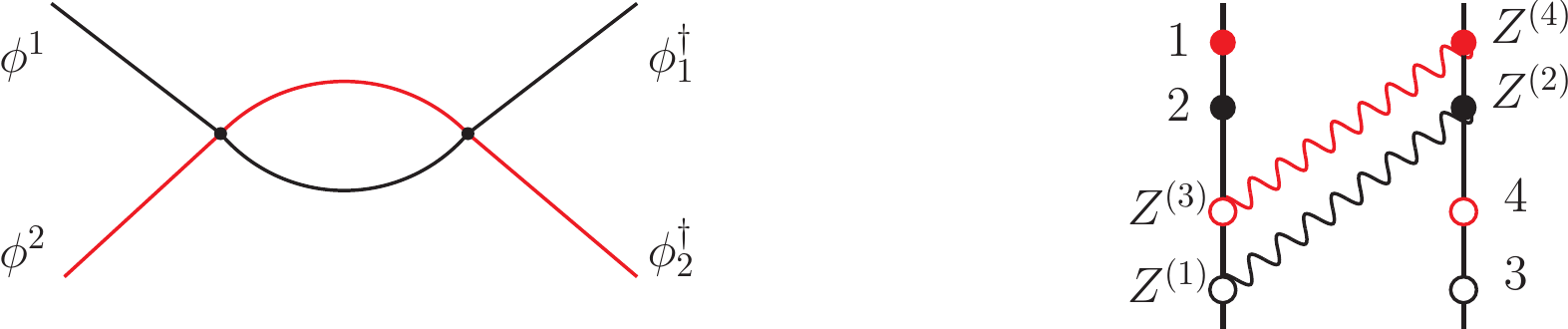}
\caption{A class of diagram leading to a double-trace UV divergence in space-time (left) and twistor space (right).}
\label{CFdiv1}
\end{figure}

Consider the 4-point 1-loop diagram given by figure~\ref{CFdiv1}, in both space-time and twistor space. Using the twistorial Feynman rules, this diagram gives 
\begin{multline}\label{primdiv1}
\tr(\sT^{\sa_1}\sT^{\sa_2})\,\tr(\sT^{\sa_3}\sT^{\sa_4}) \int_{\PT^4}\prod_{i=1}^{4}\D^{3}Z^{(i)}\,\bar{\delta}^{2}(Z^{(1)},*,Z^{(2)})\,\bar{\delta}^{2}(Z^{(3)},*,Z^{(4)}) \\
\times V_4(1,2,Z^{(3)},Z^{(1)})\,V_{4}(Z^{(4)},Z^{(2)},4,3)\,,
\end{multline}
in the planar limit, where the $Z^{(i)}$ are the propagator endpoints which are integrated over. Using \eqref{vdecomp}, the integrals in \eqref{primdiv1} can be reduced to
\be\label{primdiv2}
\int_{\PT^2}\D^3Z\, \D^3Z'\,\bar{\delta}^{1}(2,Z',*,Z)\,\bar{\delta}^{1}(3,4,*,Z')\,\cV_3(1|2,Z')\,\cV_{3}(Z|4,3)\,,
\ee
where the weights on the distributional forms can be deduced from the requirement of projective homogeneity. The object $\bar{\delta}^1$ is a $(0,1)$-distribution with support where its four arguments are projectively coplanar; for instance
\be\label{1distr}
\bar{\delta}^{1}_{-1,-1,0,-2}(2,Z',*,Z):=\int \d s\,\frac{\d t}{t}\,\d r\,r\,\bar{\delta}^{4}(Z_2+sZ'+tZ_* +rZ)\,.
\ee
Applying \eqref{vdecomp} yet again, another of the twistor integrals can be performed to leave:
\be\label{primdiv3}
\int_{\PT}\D^{3}Z\,\cV_{2}(4,3)\,\cV_{2}(1,2)\:\frac{\bar{\delta}^{1}(4,3,*,Z)\,\bar{\delta}^{2}(2,Z,1)}{(Z234)}\,,
\ee
with the weights of the distributional forms implicit.

It follows that
\be\label{primdiv4}
\frac{\bar{\delta}^{2}_{-1,-1,-2}(2,Z,1)}{(Z234)}=\frac{\bar{\delta}^{2}_{-1,-2,-1}(2,Z,1)}{(1234)}\,,
\ee
on the support of the distributional form in the numerator. This allows the final twistor integral in \eqref{primdiv3} to be performed, so this 1-loop diagram gives:
\be\label{primdiv5}
\tr(\sT^{\sa_1}\sT^{\sa_2})\,\tr(\sT^{\sa_3}\sT^{\sa_4})\,I^{(1)}(1,2|4,3)\,,
\ee
where the primitive $I^{(1)}$ is defined by
\be\label{primdiv}
I^{(1)}(1,2|4,3):=\frac{\cV_{2}(1,2)\,\cV_{2}(4,3)}{(1234)^2}\,.
\ee
Note that all dependence on the twistor $Z_*$ defining the axial gauge has dropped out, and a double pole structure (to be expected from the bubble diagram topology) emerges.

There are key differences between $I^{(1)}$ and the structure arising from analogous diagrams in planar $\cN=4$ SYM. For $\cN=4$ SYM this diagram gives an ambiguous answer in twistor space of the form `$0/0$' due to IR divergences~\cite{Brandhuber:2004yw,Bena:2004xu,Adamo:2011cb}, but \eqref{primdiv5} is finite. This is unsurprising, since FCFT should be free from IR ambiguities and the mechanisms which produced them in $\cN=4$ SYM (degenerate configurations in the \emph{fermionic} directions of $\PT$) are absent here. However, the repeated conformal invariant $(1234)^2$ in the denominator is a new structure, which does not arise in the context of $\cN=4$ SYM.

\begin{figure}[t]
\centering
\includegraphics[scale=.7]{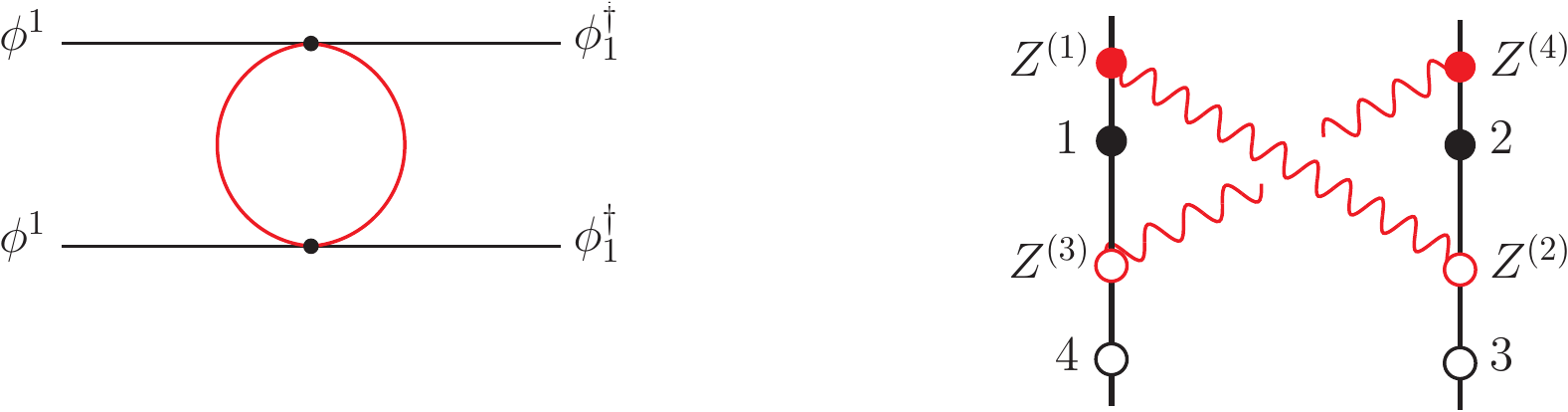}
\caption{Another class of double-trace UV divergence in space-time (left) and twistor space (right).}
\label{CFdiv2}
\end{figure}

To understand the space-time interpretation of this repeated denominator, $I^{(1)}$ can be translated to momentum space using the pairing between the cohomological expression \eqref{primdiv} and twistor momentum eigenstates. This yields:
\begin{multline}\label{primdiv6}
\int I^{(1)}(1,2|4,3)\,\prod_{i=1}^{4}\D^{3} Z_i\,\d s_i\,s_i\,\bar{\delta}^{2}\!\left(\la i|-s_i\,\lambda_i\right)\,\e^{\im s_i\,[\mu_i\,i]}=\int\frac{\d^{4}x\,\d^{4}y}{(x-y)^4}\,\e^{\im(k_1+k_2)\cdot x}\,\e^{\im(k_3+k_4)\cdot y} \\
=\delta^{4}\!\left(\sum_{i=1}^{4}k_i\right)\left(\frac{1}{\varepsilon}+\cdots\right)\,,
\end{multline}
in dimensional regularization with $d=4-2\varepsilon$. Thus, we learn that the repeated SL$(4,\C)$ invariant in the denominator of \eqref{primdiv} encodes a primitive UV-divergence in FCFT. The same structure arises from the 1-loop diagram in figure~\ref{CFdiv2}, which yields
\be\label{primdiv7}
\tr(\sT^{\sa_1}\sT^{\sa_2})\,\tr(\sT^{\sa_3}\sT^{\sa_4})\,I^{(1)}(1,4|2,3)\,.
\ee
Both classes of UV-divergence \eqref{primdiv5}, \eqref{primdiv7} are precisely what is expected from space-time considerations.

\medskip

While this gives a nice rubric for recognizing UV-divergences on twistor space (i.e., in terms of repeated conformal invariants in the denominator), the twistor approach does not immediately offer any new insights on how to actually perform loop integrations. That is, to actually extract the $\varepsilon^{-1}$ divergence, the cohomological amplitude was paired with wavefunctions and converted into a `standard' position space Feynman integral. 

Nevertheless, it may be possible to use the conformal invariance (and its breaking) encoded in twistors to derive properties of the integrals themselves \emph{before} pairing with wavefunctions. In particular, it has been shown that for finite conformal integrals arising in the context of FCFT, conformal Ward identities can be derived by probing certain singular collinear configurations~\cite{Chicherin:2017bxc}. As in the context of (super-)conformal symmetry breaking in $\cN=4$ SYM~\cite{Bullimore:2011kg}, it may be possible to reformulate these Ward identities entirely in terms of twistorial expressions, thereby gaining information about the full integral in terms of lower-loop data without having to explicitly perform any loop integrations. While such issues remain beyond the scope of this paper, they seem a promising line of enquiry for future work.

\medskip

The presence of these double trace divergences leads to the addition of counterterms in the twistor action. These are given by writing twistor vertices which generate the double trace interactions appearing in the divergences:
\begin{multline}\label{TACT}
 \alpha_1^2\,\oint_{\R^4}\d^{4}X\,\int_{(\P^1)^4}\prod_{i=1}^{4}\D\sigma_i\,\left[\mathrm{tr}\left(\phi^{1}_{1}\,\phi^{1}_{2}\right)\,\tr\left((\phi^{\dagger}_1)_3\,(\phi^{\dagger}_{1})_4\right)+\mathrm{tr}\left(\phi^{2}_{1}\,\phi^{2}_{2}\right)\,\tr\left((\phi^{\dagger}_2)_3\,(\phi^{\dagger}_{2})_4\right)\right] \\
 -\alpha_2^2\,\oint_{\R^4}\d^{4}X\,\int_{(\P^1)^4}\prod_{i=1}^{4}\D\sigma_i\,\left[\mathrm{tr}\left(\phi^{1}_{1}\,\phi^{2}_{2}\right)\,\tr\left((\phi^{\dagger}_1)_3\,(\phi^{\dagger}_{2})_4\right)+\mathrm{tr}\left(\phi^{1}_{1}\,(\phi^{\dagger}_2)_{2}\right)\,\tr\left(\phi^{2}_{3}\,(\phi^{\dagger}_1)_4\right)\right]\,,
\end{multline}
where $\alpha_1,\alpha_2$ are the induced couplings. These counterterms are invariant under the twistor gauge transformations \eqref{cfgauge}, and equal to the double-trace counterterms of FCFT in Woodhouse harmonic gauge.

There is now considerable perturbative evidence that the $\beta$-functions for the couplings $\alpha_1,\alpha_2$ have two fixed points, for which FCFT is a true (non-unitary) CFT. For $\alpha_2$, the $\beta$-function can be computed exactly~\cite{Sieg:2016vap,Grabner:2017pgm}, and vanishes for $\alpha_2^2=\xi^2$. The $\beta$-function for $\alpha_1$ has been computed up to seven loops~\cite{Grabner:2017pgm}, where it vanishes at the values
\be\label{CFT1}
\alpha_1^2=\alpha^{2}_{\pm}:=\pm\im\,\frac{\xi^2}{2}-\frac{\xi^4}{2}\mp\im\, \frac{3\,\xi^6}{4}+\xi^8\pm\im\,\frac{65\,\xi^{10}}{48}-\frac{19\,\xi^{12}}{10}+O(\xi^{14})\,.
\ee
The complexity of $\alpha^2_{\pm}$ is a consequence of the lack of unitarity; note that the two fixed points are related by $\xi^2\leftrightarrow-\xi^2$. Without loss of generality, we assume that the induced couplings lie at the CFT fixed point
\be\label{CFT2}
\alpha_1^2=\alpha^2_+\,, \qquad \alpha_2^2=\xi^2\,,
\ee
for all further calculations. Note that this conformal fixed point was obtained in the MS-bar renormalization scheme; choosing a different scheme corresponds to a finite renormalization of the coupling constant under which the conformal fixed point \eqref{CFT2} in MS-bar is mapped to a fixed point in the new scheme.


\section{Scattering amplitudes in twistor space}
\label{SCAMPS}

We are now in a position to compute general classes of cohomological scattering amplitudes for conformal fishnet theory in twistor space. The twistor formulation ensures that these cohomological answers are formulated entirely in terms of conformal invariants. The global U$(1)\times$U$(1)$ symmetry of FCFT means that non-vanishing amplitudes must have the same number of external $\phi^{1}$ and $\phi^{\dagger}_1$ fields, and similarly for $\phi^2$, $\phi^{\dagger}_2$. Consequently, scattering amplitudes can be labeled as $A_{n}(m,p)$: this is a $n$-point amplitude with $m$ external $\phi^1$ fields and $p$ external $\phi^2$ fields, where $2m+2p=n$.

The amplitudes of FCFT can be expanded in colour traces, much like amplitudes in gauge theory. However, unlike gauge theory (where there are $L+1$ trace structures at $L$ loops in perturbation theory, with only single traces in the planar limit) the amplitudes of FCFT can have multi-trace contributions at \emph{all} orders in perturbation theory, even in the large-$N$ limit. This is due to the double trace counterterms needed to ensure conformality.

Thus, a scattering amplitude of FCFT in the planar limit admits a double expansion in loops as well as traces:
\be\label{mtexpand}
A_{n}(m,p)=\delta^{4}\!\left(\sum_{i=1}^n k_i\right)\,\sum_{L=0}^{\infty} \sum_{\tau=1}^{\frac{n}{2}} \tr(\cdots)^{\tau}\,A_{n|\tau}^{L}(m,p)\,,
\ee
where $\tr(\cdots)^{\tau}$ is shorthand for $\tau$ traces over generators of SU$(N)$. The coefficient functions $A_{n|\tau}^{L}(m,p)$ are `partial amplitudes' depending only on the kinematics on space-time, or on the twistors associated with the external fields on twistor space. 

In this section, we compute the cohomological (full and partial) amplitudes for various configurations in FCFT in the planar limit at the conformal fixed point \eqref{CFT2}.


\subsection{Exact half-track amplitudes}

Consider amplitudes $A_{n}(1,\frac{n}{2}-1)$ or $A_{n}(\frac{n}{2}-1,1)$; note that $n$ must be even due to the global U$(1)\times$U$(1)$ symmetry. Since the theory is invariant under $\phi^1\rightarrow (\phi^1)^{\mathrm{T}}$, $\phi^2\rightarrow(\phi^2)^{\mathrm{T}}$, it suffices to consider the former class. In~\cite{Korchemsky:2018hnb} it was shown that $A_{4}(1,1)$ is tree-level exact; let us review the argument here. 

\begin{figure}[t]
\centering
\includegraphics[scale=.7]{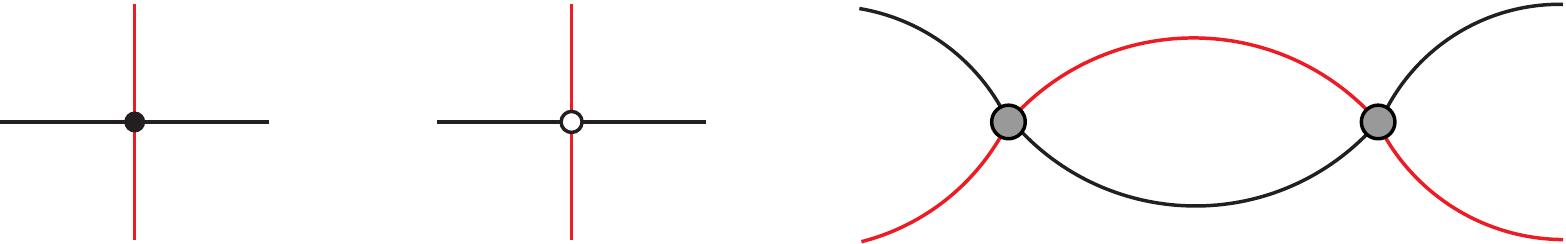}
\caption{Tree-level and 1-loop contributions to $A_{4}(1,1)$ on space-time. Black vertices stand for single trace $\xi^2$ interactions, white vertices for double trace $\alpha_2^2$ interactions and grey vertices for a combination of the two.}
\label{HTdiag1}
\end{figure}

Figure~\ref{HTdiag1} displays the tree-level contributions to $A_{4}(1,1)$ from the single trace vertex and the double trace vertices proportional to $\alpha_2^2$, as well as the 1-loop diagram which receives contributions from both of these single and double trace vertices. Therefore, $A_{4}^{0}(1,1)=A_{4|1}^{0}(1,1)+A_{4|2}^{0}(1,1)$, with
\be\label{HT4p1}
A_{4|1}^{0}(1,1)=\xi^2\,\tr(\sT^{\sa_1}\cdots\sT^{\sa_4})\,, 
\ee
\begin{equation*}
 A_{4|2}^{0}(1,1)=-\alpha_2^2\left[\tr(\sT^{\sa_1}\sT^{\sa_2})\,\tr(\sT^{\sa_3}\sT^{\sa_4})+\tr(\sT^{\sa_1}\sT^{\sa_4})\,\tr(\sT^{\sa_2}\sT^{\sa_3})\right]\,.
\end{equation*}
Due to the chiral nature of the single trace vertex, both the $\xi^2$ and $\alpha_2^2$ interactions contribute to the \emph{same} trace structure at 1-loop in the large-$N$ limit:
\be\label{HT4p2}
A_{4}^{1}(1,1)=(\xi^2-\alpha_2^2)^2\,\left[\tr(\sT^{\sa_1}\sT^{\sa_2})\,\tr(\sT^{\sa_3}\sT^{\sa_4})+\tr(\sT^{\sa_1}\sT^{\sa_4})\,\tr(\sT^{\sa_2}\sT^{\sa_3})\right]\,I\,,
\ee
where $I$ is a 1-loop integral. Evaluated at the conformal fixed point $\alpha_2^2=\xi^2$, the 1-loop amplitude vanishes, as do all higher-loop corrections by the same mechanism~\cite{Korchemsky:2018hnb}.

\begin{figure}[t]
\centering
\includegraphics[scale=.7]{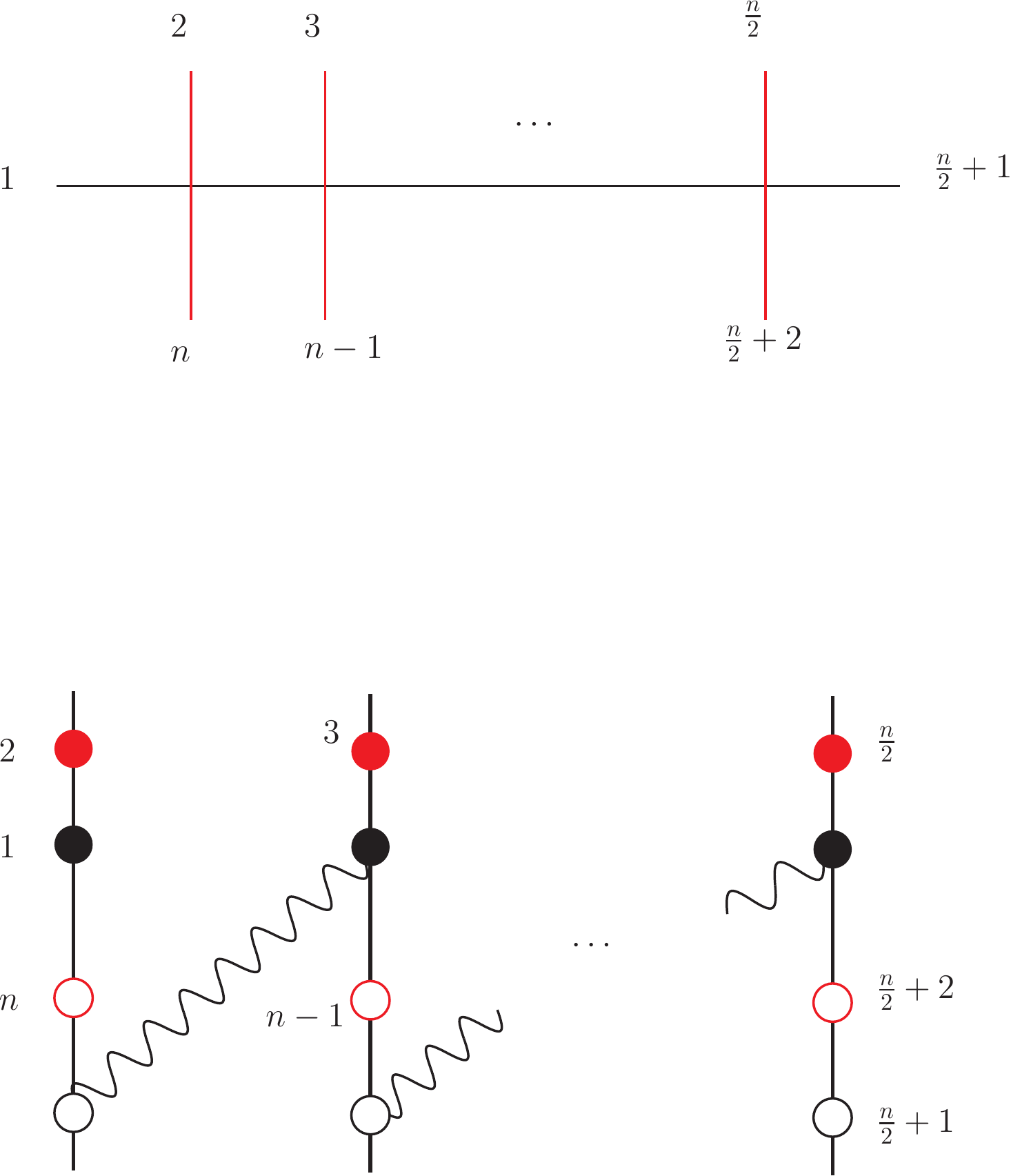}
\caption{The structure of a half-track amplitude in space-time (top) and twistor space (bottom).}
\label{HTdiag2}
\end{figure}

This argument generalizes to all amplitudes of the form $A_{n}(1,\frac{n}{2}-1)$. At tree-level, these are represented by half-track diagrams -- shown in space-time and twistor space in figure~\ref{HTdiag2} -- where vertices are given by single trace $\xi^2$ interactions or double trace $\alpha_2^2$ interactions. Quantum corrections entail the insertion of the loops appearing in figure~\ref{HTdiag1}, which are always proportional to $(\xi^2-\alpha_2^2)^2$ and vanish at the conformal fixed point. Thus, all half-track amplitudes are tree-level exact.  

The trace structure of the tree-level contribution is dictated by the location and number of double trace vertices appearing in the half-track diagram. Indeed, if $k$ of the $\frac{n}{2}-1$ total vertices are double trace (proportional to $\alpha_2^2$) then the corresponding contribution will have $k+1$ traces. It is easy to work out the particular trace structure for a given vertex placement, but in any case all of the trace structures share the same kinematic structure. In particular, at the conformal fixed point the half-track amplitude is always
\be\label{HT1}
 A_{n}\left(1,\frac{n}{2}-1\right)=\xi^{n-2}\,\left(\mbox{traces}\right)\,K_{n}^{0}\left(1,\frac{n}{2}-1\right)\,,
\ee
where $K_{n}^{0}(1,\frac{n}{2}-1)$ is the kinematic function associated with the diagram~\ref{HTdiag2}.

In momentum space this kinematic function is simply a product of scalar propagators. But in twistor space, it is given by the cohomological amplitude that follows from evaluating the twistor half-track diagram in the bottom of figure~\ref{HTdiag2}. Using the twistor propagator in axial gauge, we find:
\begin{multline}\label{HT2}
K_{n}^{0}\left(1,\frac{n}{2}-1\right)=\frac{\cV_{3}(n|1,2)\,\cV_{3}(\frac{n}{2}|\frac{n}{2}+1,\frac{n}{2}+2)}{(1\,2\,3\,n-1)\,(\frac{n}{2}-1\,\frac{n}{2}+3\,\frac{n}{2}+1\,\frac{n}{2}+2)} \\
\times\,\prod_{i=3}^{\frac{n}{2}-2}\frac{\cV_{2}(i,n-i+2)\,\cV_{2}(i+1,n-i+1)}{(i\,n-i+2\,i+1\,n-i+1)}\,.
\end{multline}
As expected, this result is built entirely from SL$(4,\C)$ invariants and manifestly independent of the twistor $Z_*$ that defines the axial gauge condition. Equivalent representations can be constructed by making alternative decompositions of the 4-point vertices using \eqref{vdecomp}.


\subsection{Four-point single colour amplitude}

In contrast to half-track amplitudes, the `single colour' four-point amplitude $A_{4}(2,0)$ has non-vanishing quantum corrections at each loop order~\cite{Korchemsky:2018hnb}. Many properties of $A_{4}(2,0)$ were studied in~\cite{Korchemsky:2018hnb} at both weak and strong coupling, but it is particularly illustrative to study the cancellation of UV-divergences between diagrams at each loop order in twistor space.

\begin{figure}[t]
\centering
\includegraphics[scale=.7]{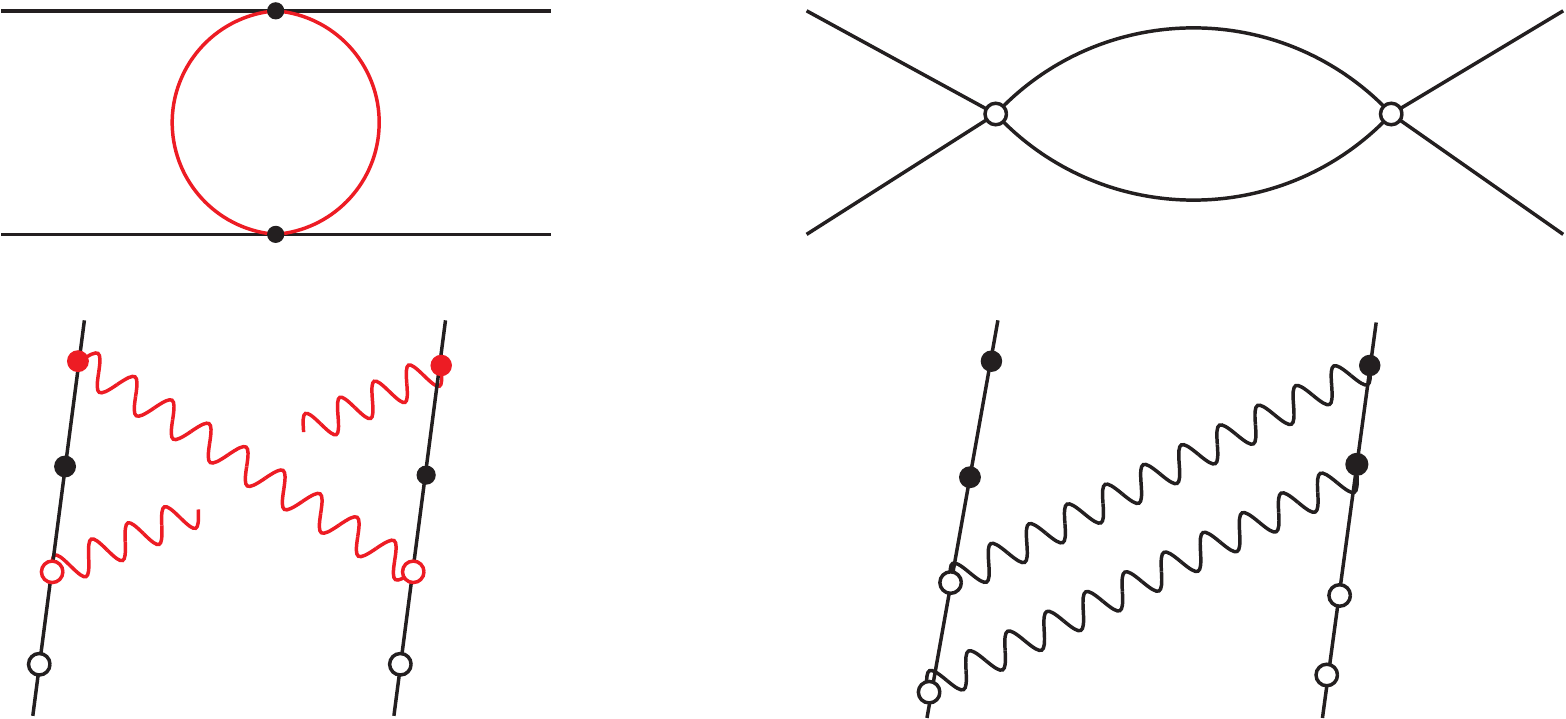}
\caption{The two types of 1-loop diagram contributing to $A_4(2,0)$ on space-time (upper) and twistor space (lower).}
\label{SC1loopd}
\end{figure}

In the planar limit, $A_4(2,0)$ is a double trace structure $\tr(\sT^{\sa_1}\sT^{\sa_2})\,\tr(\sT^{\sa_3}\sT^{\sa_4})$ to all loop orders; the tree-level contribution is given by the first term in the $\alpha_1^2$ interaction of \eqref{TACT}. On twistor space, this is simply
\be\label{SCtree}
A^{0}_{4|2}(2,0)=2\im\,\xi^2\,V_{4}(1,2,3,4)\,,
\ee
where the overall factor is $4\alpha_1^2$ evaluated at the conformal fixed point. At 1-loop, there are two classes of diagram that contribute, as shown in figure~\ref{SC1loopd}. The first of these we have already evaluated in \eqref{primdiv7}, and the second is easily computed using the same techniques, with the result:
\be\label{SC1loop}
A^{1}_{4|2}(2,0)=\xi^4\,I(1,4|2,3)+\xi^{4}\,I(2,4|1,3)+8\alpha_{+}^{4}\,I(1,2|3,4)\,,
\ee
with $I$ defined by \eqref{primdiv}.

Clearly, each term in \eqref{SC1loop} is UV-divergent: $I$ encodes a simple pole (in dimensional regularization) which is independent of its arguments, due to the presence of squared conformal invariants in the denominator. However, when the value of the conformal fixed point $\alpha_+^2=\im\xi^2/2$ is inserted,
\be\label{SC1loop*}
A^{1}_{4|2}(2,0)=\xi^4\,\left[I(1,4|2,3)+I(2,4|1,3)-2\,I(1,2|3,4)\right]\,.
\ee
Since the singular parts of each term in the brackets are independent of the arguments of $I$, all of the divergences cancel, leaving a finite remainder. The precise form of this finite remainder (after pairing with external wavefunctions and integrating) depends on the renormalization scheme, which we assume to be MS-bar. 

\begin{figure}[t]
\centering
\includegraphics[scale=.7]{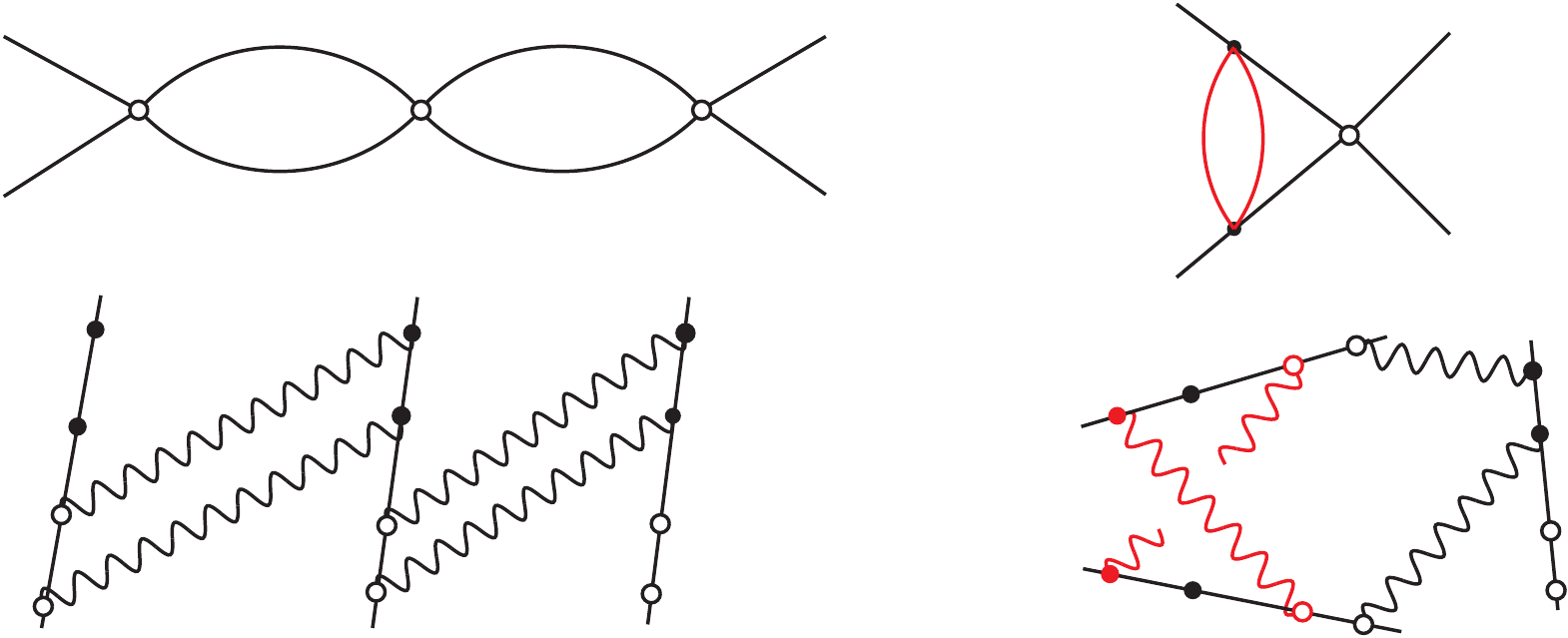}
\caption{The two types of 2-loop diagram contributing to $A_4(2,0)$ on space-time (upper) and twistor space (lower).}
\label{SC2loopd}
\end{figure}

To see what happens at two loops, we must confront a new situation. There are two classes of diagrams which contribute to $A_{4|2}^{2}(2,0)$, as illustrated in space-time and twistor space in figure~\ref{SC2loopd}. In both cases, we see that the twistor diagrams involve a vertex/line with less than two external insertions; this scenario never occurs in the protected half-track amplitudes. Let us consider the first class of diagram; using the usual twistor machinery leads us to an expression of the form:
\be\label{SC2l1}
\cV_{2}(1,2)\,\cV(3,4) \oint\frac{\D^{3}A\wedge\D^{3}B}{(12AB)\,(AB34)}\,\bar{\delta}^{1}(1,2,*,A)\,\bar{\delta}^{1}(B,*,3,4)\,.
\ee
At first, one might be tempted to conclude that there are no UV-divergences encoded in this diagram, since the two conformal invariants in the denominator are different. However, the distributional forms in the numerator can still be reduced to non-projective distributions as, so this conclusion is premature.

Indeed, it is an easy exercise to confirm that
\be\label{SC2l2}
\bar{\delta}^{1}_{-1,-1,0,-2}(1,2,*,A)=-\frac{(12*B)}{(12AB)}\,\bar{\delta}(12*A)\,,
\ee
where $\bar{\delta}(z)=\dbar z^{-1}$ is the standard $(0,1)$-distribution of weight $-1$. This identity -- applied to both distributional forms inside of the integral -- enables \eqref{SC2l1} to be rewritten as
\be\label{SC2l3}
\cV_{2}(1,2)\,\cV(3,4) \oint\frac{\D^{3}A\wedge\D^{3}B}{(12AB)^2\,(AB34)^2}\,\bar{\delta}(12*A)\,\bar{\delta}(B*34)\,(12*B)\,(34*A)\,.
\ee
There are now two repeated conformal invariants in the denominator, leading us to expect that this cohomological expression encodes a \emph{quadratic} UV-divergence on space-time. Indeed, when \eqref{SC2l3} is paired with momentum eigenstate wavefunctions, it returns the space-time integral
\be\label{SC2l4}
\int \frac{\d^{4}x\,\d^{4}y\,\d^{4}z}{(x-z)^4\,(z-y)^4}\,\e^{\im (k_1+k_2)\cdot x}\,\e^{\im (k_3+k_4)\cdot y}=\delta^{4}\!\left(\sum_{i=1}^{4}k_i\right)\left(\frac{1}{\varepsilon^2}+\cdots\right)\,,
\ee
in dimensional regularization.

Similarly, the second class of diagram in figure~\ref{SC2loopd} can be evaluated to
\be\label{SC2l5}
\cV_{2}(3,4)\oint\frac{\D^{3}A\wedge\D^{3}B\,\D^{3}C\wedge\D^{3}D}{(1B34)^2\,(2D34)^2}\bar{\delta}^{3}(1,A)\,\bar{\delta}^{3}(2,C)\,\bar{\delta}(1B*D)\,\bar{\delta}(2B*D)\,(1*34)(2*34)\,,
\ee
which features the same denominator structure leading to a quadratic UV-divergence. Adding together all of the diagrams with the appropriate symmetry factors yields
\be\label{SC2l6}
16\,\alpha_+^{6}\,(\mbox{diag. 1})+2\,\xi^{4}\,\alpha_+^{2}\,(\mbox{diag. 2})+2\,\xi^{4}\,\alpha_+^{2}\,(\mbox{diag. 2})_{(1,2)\leftrightarrow(3,4)}\,.
\ee
At the conformal fixed point value for $\alpha_+^2$, this has precisely the structure required to cancel the quadratic UV-divergences between all three terms. 

At general loop order, similar mechanisms will always be at play. Any $L$-loop diagram in twistor space contributing to $A_4(2,0)$ will contain $L$ repeated conformal invariants in its denominator, and all such diagrams will be combined in a fashion which cancels these divergences at the conformal fixed point of the theory. The value of the finite remainder depends on the renormalization scheme used to perform the integrals after pairing with external wavefunctions.

We also observe a generic feature of FCFT Feynman diagrams in twistor space: any diagram with fewer than two external insertions on one of its vertices will not be explicitly independent of $Z_*$. Nevertheless, such contributions to scattering amplitudes remain independent of the choice of axial gauge. This follows because such diagrams are always holomorphic and homogeneous of degree zero with respect to $Z_*$.


\subsection{Snowflake amplitudes}

\begin{figure}[t]
\centering
\includegraphics[scale=.7]{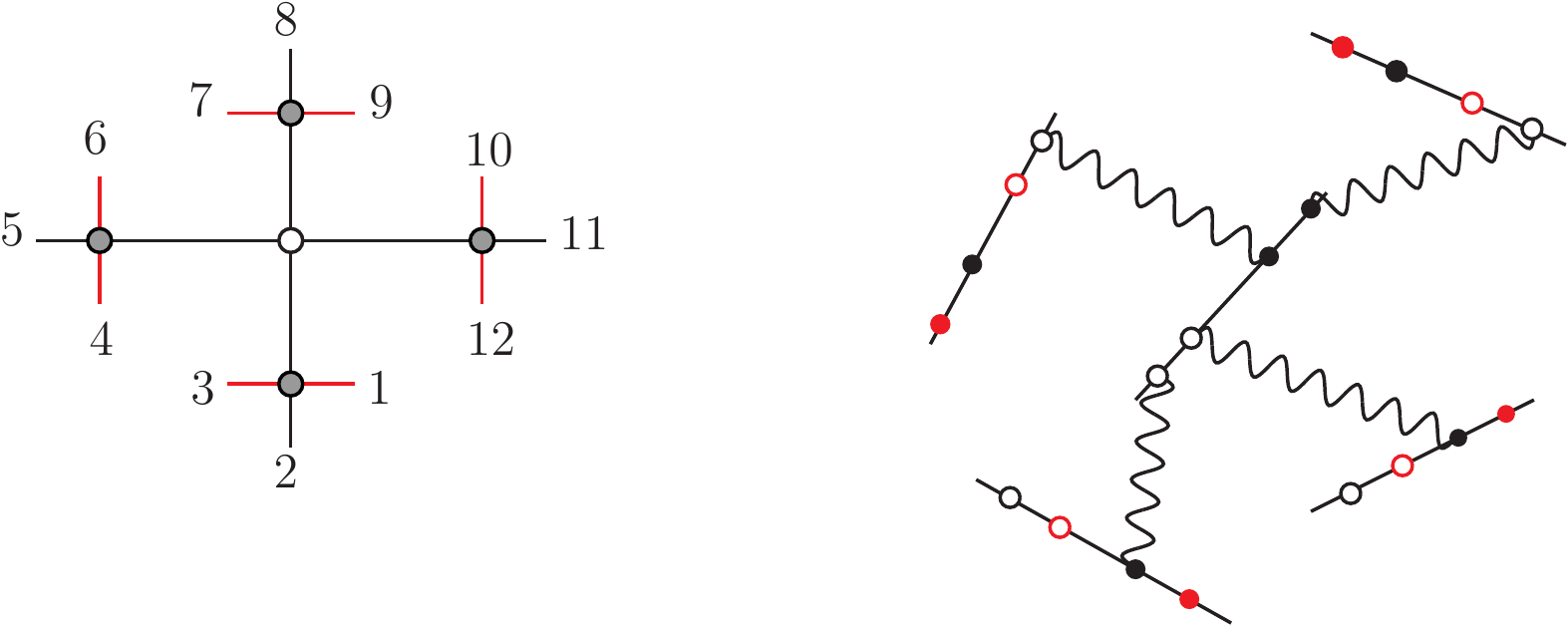}
\caption{Tree-level snowflake amplitude in space-time (left) and twistor space (right).}
\label{SFTree}
\end{figure}

Certain higher-point amplitudes in FCFT are controlled by structures inherited from the four-point single colour and half-track amplitudes. A particularly illustrative example is given by the amplitude $A_{12}(2,4)$, which we refer to as the `snowflake' due to the appearance of its space-time Feynman diagrams in the planar limit. At tree-level, this involves a single insertion of the double trace $\alpha_1^2$ interaction, linked to four insertions of single trace $\xi^2$ or double trace $\alpha_2^2$ interactions, as depicted in figure~\ref{SFTree}.  

The trace structure of the amplitude depends on where any $\alpha_2^2$ vertices appear in the snowflake diagram. If the only double trace vertex is the one appearing in the middle of the diagram, then the overall structure is double trace, with the grouping of generators determined by the U$(1)\times$U$(1)$ quantum numbers. In the space-time diagram of figure~\ref{SFTree}, suppose we number the external particles in the clockwise fashion shown in the figure. If legs $2$ and $11$ correspond to $\phi^{\dagger}_1$ fields, then the double trace structure is:
\be\label{SFt1}
\tr(\sT^{\sa_1}\sT^{\sa_2}\sT^{\sa_3}\sT^{\sa_{10}}\sT^{\sa_{11}}\sT^{\sa_{12}})\,\tr(\sT^{\sa_4}\sT^{\sa_5}\sT^{\sa_6}\sT^{\sa_{7}}\sT^{\sa_{8}}\sT^{\sa_{9}})\,.
\ee
When one of the peripheral vertices is an $\alpha^2_2$ interaction, a triple trace structure results, decomposed around the insertion. For instance, if a double trace interaction is inserted at the bottom-most vertex in figure~\ref{SFTree}, then the triple trace structure is
\be\label{SFt2}
\big[ \tr(\sT^{\sa_1}\sT^{\sa_{10}}\sT^{\sa_{11}}\sT^{\sa_{12}})\,\tr(\sT^{\sa_2}\sT^{\sa_3})+\tr(\sT^{\sa_3}\sT^{\sa_{10}}\sT^{\sa_{11}}\sT^{\sa_{12}})\,\tr(\sT^{\sa_2}\sT^{\sa_1})\big] \tr(\sT^{\sa_4}\cdots\sT^{\sa_{9}})\,.
\ee
Overall, the tree-level amplitude has a trace decomposition:
\be\label{SFtraces}
A^{0}_{12}(2,4)=\sum_{\tau=2}^{6} \tr(\cdots)^{\tau}\,A^{0}_{12|\tau}(2,4)\,,
\ee
with the maximal six-trace contribution arising when all five vertices of the snowflake diagram are double trace.

In twistor space, the cohomological amplitude associated with the snowflake is:
\begin{multline}\label{STt3}
\int \prod_{i=1}^{8}\D^{3}Z^{(i)} V_{4}(1,2,3,Z^{(1)})\cdots V_{4}(10,11,12,Z^{(4)})\,V_{4}(Z^{(5)},Z^{(6)},Z^{(7)},Z^{(8)}) \\
\times\bar{\delta}^{2}(Z^{(1)},*,Z^{(5)})\,\cdots\,\bar{\delta}^{2}(Z^{(4)},*,Z^{(8)})\,.
\end{multline}
Applying \eqref{vdecomp} and performing the maximum number of integrations leaves:
\begin{multline}\label{STt4}
A^{0}_{12|\tau}(2,4)=\pm\im\frac{\xi^{10}}{2}\:\cV_{3}(2|3,1)\cdots \cV_{3}(11|12,10)\\
\oint \frac{\D^{3}A\wedge\D^{3}B}{(12\,10\,A\,B)\,(9\,7\,A\,B)}\,\bar{\delta}^{1}(3,1,*,A)\wedge\bar{\delta}^{1}(6,4,*,B)\,,
\end{multline}
at the conformal fixed point, with the overall sign determined by whether there are an even ($+$) or odd ($-$) number of $\alpha_2^2$ insertions. The distributional forms in the integrand can be further reduced using identity \eqref{SC2l2}, giving an equivalent expression
\begin{multline}\label{STt4*}
 A^{0}_{12|\tau}(2,4)=\pm\im\frac{\xi^{10}}{2}\:\cV_{3}(2|3,1)\cdots \cV_{3}(11|12,10)\\
\oint \frac{\D^{3}A\wedge\D^{3}B}{(12\,10\,A\,B)\,(9\,7\,A\,B)}\,\frac{(3\,1\,*\,B)\,(6\,4\,*\,A)}{(3\,1\,A\,B)\,(6\,4\,A\,B)}\,\bar{\delta}(31*A)\,\bar{\delta}(64*B)\,.
\end{multline}
Since there are no UV-divergences (i.e., no repeated denominators), there is no particular advantage to using the representation \eqref{STt4*} as opposed to \eqref{STt4} besides the cosmetic symmetry of the denominator.

\medskip

\begin{figure}[t]
\centering
\includegraphics[scale=.7]{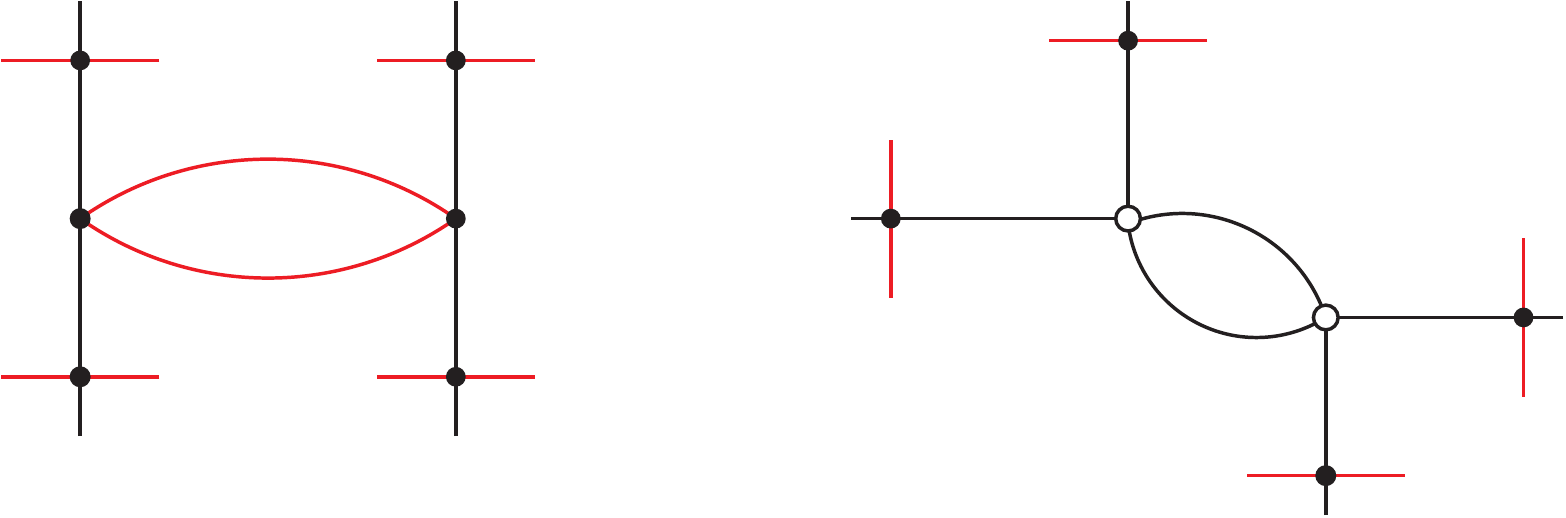}
\caption{One-loop diagrams contributing to the double trace snowflake amplitude.}
\label{SFLoop}
\end{figure}

We briefly discuss the structure of quantum corrections to the snowflake amplitude. By the arguments of~\cite{Korchemsky:2018hnb}, it follows that there are no 1-loop corrections to the outer four-point interactions (in the planar limit and at the conformal fixed point). For simplicity, assume that all four of these outer interactions are single trace. On space-time, the central 4-point interaction can be corrected by the 1-loop diagrams shown in figure~\ref{SFLoop}; both contribute with the same overall double trace structure \eqref{SFt1}. 

The 1-loop kinematic contribution from all relevant diagrams of this form is given by
\be\label{SFl1}
\xi^{4}\,F(k_{123789})+\xi^{4}\,F(k_{10\,11\,12\,789})-2\xi^{4}\,F(k_{123\,10\,11\,12})\,,
\ee
where the coefficient of the final term is equivalent to $8\alpha_+^{2}$ evaluated at the fixed point \eqref{CFT2}, $k_{i\cdots j}:=k_i+\cdots+k_j$, and $F$ is the 1-loop scalar integral
\be\label{SFl2}
F(k_{i\cdots j}):=\int\frac{\d^{4-2\varepsilon}\ell}{(2\,\pi)^{4-2\varepsilon}}\,\frac{1}{\ell^2\,(k_{i\cdots j}-\ell)^2}\,.
\ee
Term-by-term, \eqref{SFl1} is UV-divergent, but it is straightforward to see that the combination of all three terms is finite. This is the same mechanism that removes all divergences from the four-point single colour amplitudes (cf., \cite{Korchemsky:2018hnb}) at the conformal fixed point.

On twistor space, the diagram corresponding to the first term in \eqref{SFl1} evaluates to:
\be\label{SFl3}
\xi^{12}\,\cV_{3}(2|3,1)\cdots \cV_{3}(11|12,10)\,\cF[3,1|9,7][6,4|12,10]\,,
\ee
where the integral $\cF$ is defined by:
\begin{multline}\label{SFl4}
\cF[a,b|c,d][i,j|k,l]:=\oint \frac{\D^{3}A\wedge\D^{3}B\:\D^{3}C\wedge\D^{3}D}{(ABCD)^2}\:\bar{\delta}^{1}(a,b,*,A)\,\bar{\delta}^{1}(c,d,*,B)\\
\bar{\delta}^{1}(i,j,*,C)\,\bar{\delta}^{1}(k,l,*,D)\,.
\end{multline}
The squared conformal invariant in the denominator is indicative of a linear UV-divergence, and pairing with external twistor wavefunctions confirms that there is a simple pole (in dimensional regularization) which is independent of the arguments of $\cF$. Upon evaluating all 1-loop diagrams in twistor space contributing to the snowflake amplitude, one finds
\begin{multline}\label{SFl5}
 A^{1}_{12|2}(2,4)=\xi^{12}\,\cV_{3}(2|3,1)\cdots \cV_{3}(11|12,10)\bigg(\cF[3,1|9,7][6,4|12,10] \\
 +\cF[3,1|6,4][9,7|12,10]-2\,\cF[3,1|12,10][6,4|9,7]\bigg)\,,
\end{multline}
at the conformal fixed point. Sure enough, since the divergent part of each term is independent of its arguments, all UV-divergences cancel from the 1-loop result.


\subsection{Fishnet diagrams}

As a final example, consider planar fishnet diagrams where all vertices are given by the single trace interaction. These diagrams contribute to the single trace component of the amplitude in the planar limit, and combine many of the features of twistor amplitudes observed in the preceding examples. Every twistor space vertex involved in a fishnet diagram has no more than two external field insertions, so vertices of a general fishnet diagram fall into one of three sets: corner (two external field insertions), edge (one external insertion) and interior (no external insertions) vertices.

\begin{figure}[t]
\centering
\includegraphics[scale=.7]{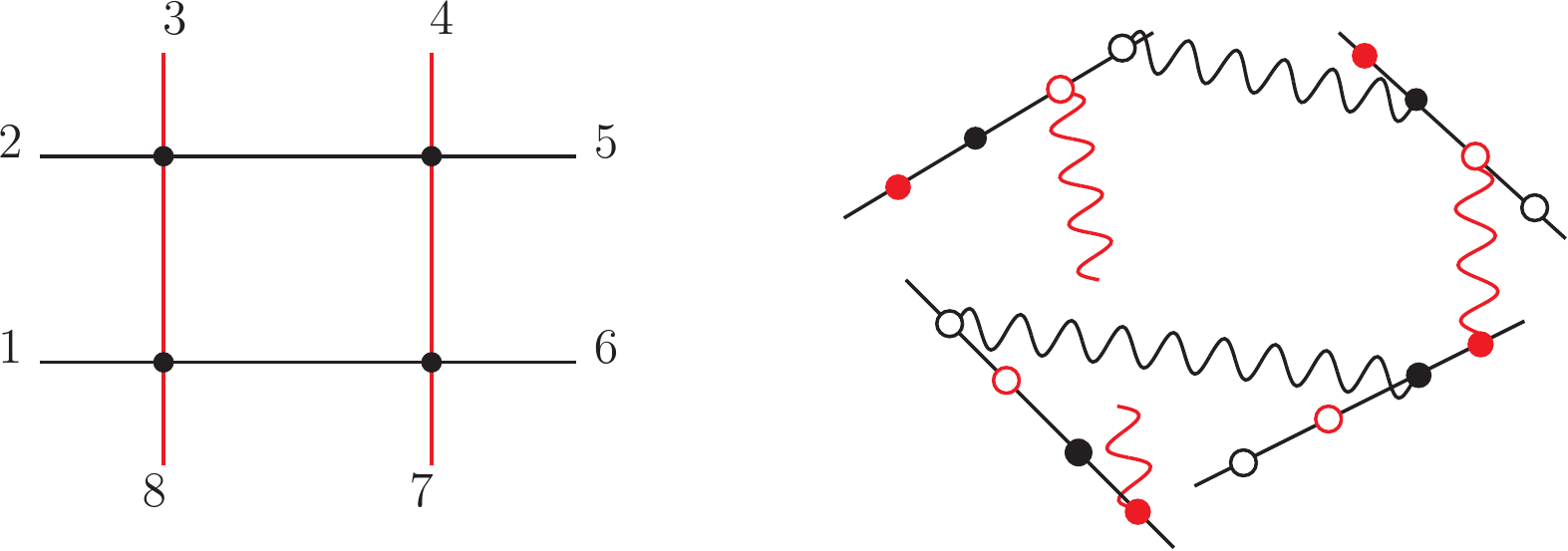}
\caption{The 1-loop fishnet diagram on space-time (left) and twistor space (right).}
\label{FN1loop}
\end{figure}

The 1-loop case illustrated in figure~\ref{FN1loop} is straightforward to evaluate:
\be\label{FNl1}
A_{8|1}^{1}(2,2)=\frac{\cV_{2}(8,1)\,\cV_{2}(2,3)\,\cV_{2}(4,5)\,\cV_{2}(6,7)}{(8123)\,(2345)\,(4567)\,(6781)}\,,
\ee
which is manifestly conformally invariant and free from UV- and IR-divergences. Pairing with external wavefunctions gives the standard conformal box integral, which can be evaluated in terms of dilogarithms~\cite{Usyukina:1992jd}. So if $V_{4}(i,j,\bullet,\bullet)$ and $V_{4}(k,\bullet,\bullet,l)$ are two corner vertices of a fishnet diagram connected by a propagator, their contraction will produce a factor of
\be\label{FNccon}
\frac{\cV_{2}(i,j)\,\cV_{2}(k,l)}{(ijkl)}\,,
\ee
in the evaluation of the diagram.

Next, consider a propagator contraction between a corner vertex $V_{4}(i,j,\bullet,\bullet)$ and an edge vertex $V_{4}(k,\bullet,\bullet,\bullet)$. This contraction can be evaluated to
\be\label{FNcecon}
\cV_{2}(i,j)\oint \frac{\D^{3}A\wedge\D^{3}B}{(ijkB)}\,\bar{\delta}^{3}(k,A)\,\bar{\delta}^{3}(\bullet,B)\,\bar{\delta}^{2}(k,\bullet,\bullet)\,,
\ee
where the remaining $\bullet$ entries are contracted with propagators to some other part of the diagram. All remaining contractions (between edge and interior vertices) are then dictated by the remaining distributional forms. 

For a generic fishnet diagram $\Gamma$ contributing to $A_{n}(m,k)$, each vertex can be labeled by a pair $(i,j)$ for $i\in\{1,\ldots,m\}$ and $j\in\{1,\ldots,k\}$. Label the sets of corner, edge and interior vertices by $\mathcal{C}$, $\mathcal{E}$ and $\mathcal{I}$, respectively. The cohomological amplitude associated with $\Gamma$ takes the form
\be\label{FN1}
\oint \prod_{(i,j)}\D^{3}A_{ij}\wedge\D^{3}B_{ij}\:\mathcal{B}(\mathcal{C}\cup\mathcal{E})\,\prod_{(i,j)\in\mathcal{E}\cup\mathcal{I}}\bar{\delta}^{2}(A_{ij},*,B_{(i+1)j})\,\bar{\delta}^{2}(A_{i(j+1)},*,B_{ij})\,,
\ee
where $\mathcal{B}(\mathcal{C}\cup\mathcal{E})$ is a contribution depending on the boundary topology of the fishnet diagram that encodes all dependence on external field insertions. This $\mathcal{B}$ can be determined directly from the rules described above.

\begin{figure}[t]
\centering
\includegraphics[scale=.7]{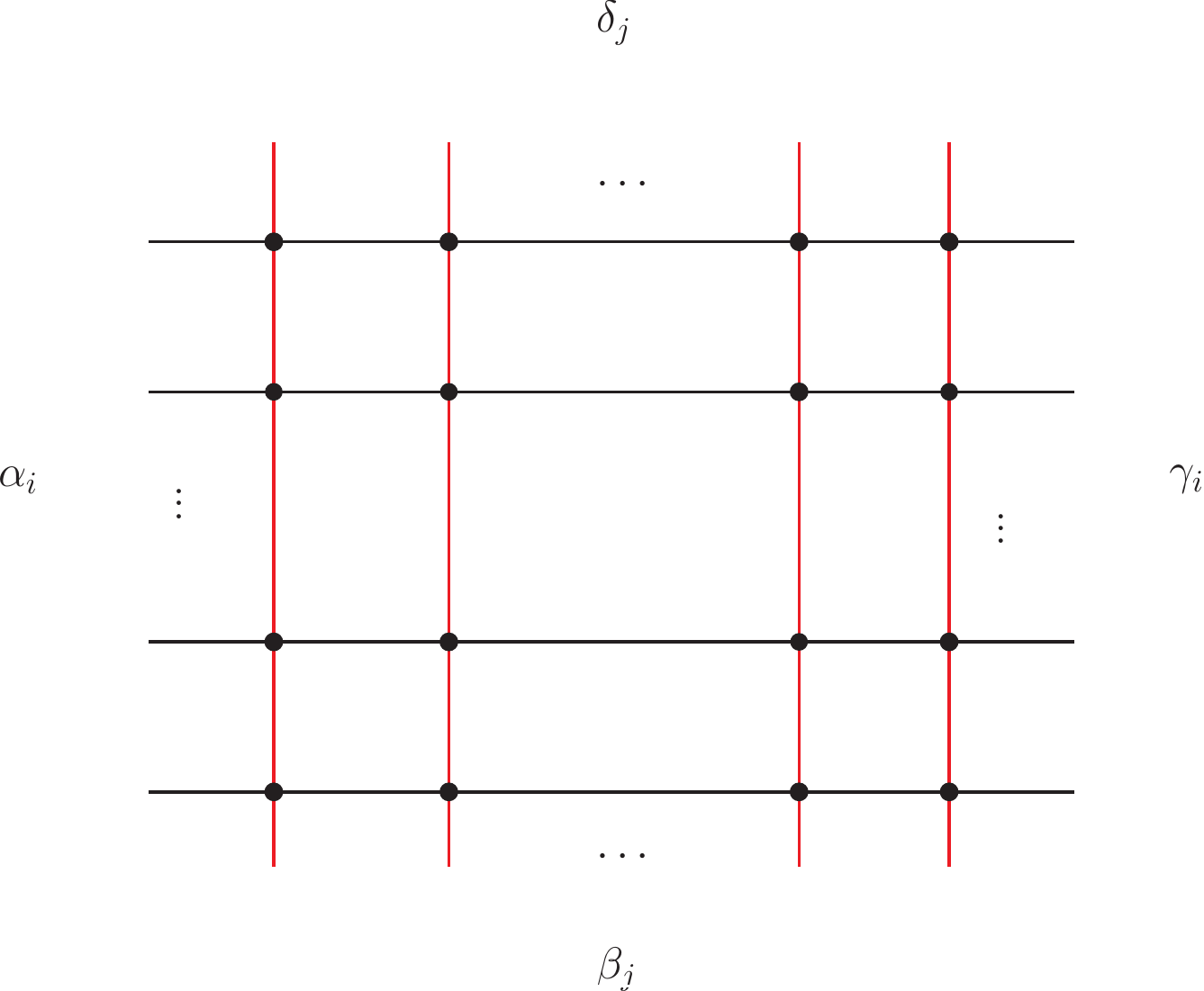}
\caption{The $m\times k$ rectangular fishnet diagram with external leg twistor labels.}
\label{FNmult}
\end{figure}

To be completely precise, let's consider an example. Suppose that $\Gamma$ is a $m\times k$ rectangular fishnet diagram (for $m,k>2$), with $(m-1)(k-1)$ loops. Label the external points on the columns $(i,1)$ and $(i,k)$ by twistors $\alpha_i$ and $\gamma_i$, respectively. Label the external points on rows $(1,j)$ and $(m,j)$ with twistors $\beta_j$ and $\delta_j$, respectively. This labeling scheme is shown in figure~\ref{FNmult}. Then the boundary contribution to \eqref{FN1} is given by
\begin{multline}\label{FNrect}
 \mathcal{B}^{\mathrm{rect}}_{m\times k}=\frac{\bar{\delta}^{3}(\alpha_1,A_{11})\,\bar{\delta}^{3}(\beta_1,B_{11})}{(\alpha_1\beta_1 A_{12}\beta_2)\,(\alpha_1\beta_1\alpha_2 B_{21})}\:\:\frac{\bar{\delta}^{3}(\alpha_m,A_{m1})\,\bar{\delta}^{3}(\delta_1,B_{m1})}{(\alpha_m\delta_1 A_{m2}\delta_2)\,(\alpha_m\delta_1\alpha_{m-1} B_{(m-1)1})} \\
 \frac{\bar{\delta}^{3}(\gamma_m,A_{mk})\,\bar{\delta}^{3}(\delta_k,B_{mk})}{(A_{m(k-1)}\delta_{k-1}\gamma_m \delta_k)\,(\gamma_{m-1}B_{(m-1)k}\gamma_m \delta_k)}\:\:\frac{\bar{\delta}^{3}(\gamma_1,A_{1k})\,\bar{\delta}^{3}(\beta_k,B_{1k})}{(A_{1(k-1)}\beta_{k-1} \gamma_{1}\beta_{k})\,(\gamma_1 \beta_k \gamma_2 B_{2k})} \\
 \prod_{i=2}^{m-2}\frac{\bar{\delta}^{3}(\alpha_{i},A_{i1})\,\bar{\delta}^{3}(\gamma_{i},A_{ik})}{(\alpha_i B_{i1}\alpha_{i+1}B_{(i+1)1})\,(\gamma_i B_{ik}\gamma_{i+1} B_{(i+1)k})}\, \prod_{j=2}^{k-2} \frac{\bar{\delta}^{3}(\beta_j,B_{1j})\,\bar{\delta}^{3}(\delta_j,B_{mj})}{(A_{1j}\beta_j A_{1(j+1)}\beta_{j+1})\,(A_{mj}\delta_j A_{m(j+1)}\delta_{j+1})}\,.
\end{multline}
Here, all projective delta functions have weight $\bar{\delta}^{3}_{0,-4}(\cdot,\cdot)$, ensuring that $\mathcal{B}^{\mathrm{rect}}_{m\times k}$ has the appropriate projective and form weights to give a well-defined cohomological amplitude upon being inserted into \eqref{FN1}.

\acknowledgments

We thank Gregory Korchemsky and Matthias Wilhelm for helpful comments on a draft, as well as Jake Bourjaily, Amit Sever and David Skinner for interesting discussions. TA is supported by an Imperial College Junior Research Fellowship.

\bibliography{CFTbib}

\providecommand{\href}[2]{#2}\begingroup\raggedright\begin{thebibliography}{10}

\bibitem{Beisert:2010jr}
N.~Beisert et~al., \emph{{Review of AdS/CFT Integrability: An Overview}},
  \href{http://dx.doi.org/10.1007/s11005-011-0529-2}{\emph{Lett. Math. Phys.}
  {\bfseries 99} (2012) 3--32},
  [\href{https://arxiv.org/abs/1012.3982}{{\ttfamily 1012.3982}}].

\bibitem{Elvang:2013cua}
H.~Elvang and Y.-t. Huang, \emph{{Scattering Amplitudes}},
  \href{https://arxiv.org/abs/1308.1697}{{\ttfamily 1308.1697}}.

\bibitem{Gurdogan:2015csr}
O.~Gürdoğan and V.~Kazakov, \emph{{New Integrable 4D Quantum Field Theories
  from Strongly Deformed Planar $\mathcal N = $ 4 Supersymmetric Yang-Mills
  Theory}}, \href{http://dx.doi.org/10.1103/PhysRevLett.117.201602,
  10.1103/PhysRevLett.117.259903}{\emph{Phys. Rev. Lett.} {\bfseries 117}
  (2016) 201602}, [\href{https://arxiv.org/abs/1512.06704}{{\ttfamily
  1512.06704}}].

\bibitem{Leigh:1995ep}
R.~G. Leigh and M.~J. Strassler, \emph{{Exactly marginal operators and duality
  in four-dimensional N=1 supersymmetric gauge theory}},
  \href{http://dx.doi.org/10.1016/0550-3213(95)00261-P}{\emph{Nucl. Phys.}
  {\bfseries B447} (1995) 95--136},
  [\href{https://arxiv.org/abs/hep-th/9503121}{{\ttfamily hep-th/9503121}}].

\bibitem{Lunin:2005jy}
O.~Lunin and J.~M. Maldacena, \emph{{Deforming field theories with U(1) x U(1)
  global symmetry and their gravity duals}},
  \href{http://dx.doi.org/10.1088/1126-6708/2005/05/033}{\emph{JHEP} {\bfseries
  05} (2005) 033}, [\href{https://arxiv.org/abs/hep-th/0502086}{{\ttfamily
  hep-th/0502086}}].

\bibitem{Frolov:2005dj}
S.~Frolov, \emph{{Lax pair for strings in Lunin-Maldacena background}},
  \href{http://dx.doi.org/10.1088/1126-6708/2005/05/069}{\emph{JHEP} {\bfseries
  05} (2005) 069}, [\href{https://arxiv.org/abs/hep-th/0503201}{{\ttfamily
  hep-th/0503201}}].

\bibitem{Frolov:2005ty}
S.~A. Frolov, R.~Roiban and A.~A. Tseytlin, \emph{{Gauge-string duality for
  superconformal deformations of N=4 super Yang-Mills theory}},
  \href{http://dx.doi.org/10.1088/1126-6708/2005/07/045}{\emph{JHEP} {\bfseries
  07} (2005) 045}, [\href{https://arxiv.org/abs/hep-th/0503192}{{\ttfamily
  hep-th/0503192}}].

\bibitem{Caetano:2016ydc}
J.~Caetano, O.~Gürdoğan and V.~Kazakov, \emph{{Chiral limit of $ \mathcal{N}
  $ = 4 SYM and ABJM and integrable Feynman graphs}},
  \href{http://dx.doi.org/10.1007/JHEP03(2018)077}{\emph{JHEP} {\bfseries 03}
  (2018) 077}, [\href{https://arxiv.org/abs/1612.05895}{{\ttfamily
  1612.05895}}].

\bibitem{Kazakov:2018gcy}
V.~Kazakov, E.~Olivucci and M.~Preti, \emph{{Generalized fishnets and exact
  four-point correlators in chiral CFT$_{4}$}},
  \href{http://dx.doi.org/10.1007/JHEP06(2019)078}{\emph{JHEP} {\bfseries 06}
  (2019) 078}, [\href{https://arxiv.org/abs/1901.00011}{{\ttfamily
  1901.00011}}].

\bibitem{Fokken:2013aea}
J.~Fokken, C.~Sieg and M.~Wilhelm, \emph{{Non-conformality of ${{\gamma
  }_{i}}$-deformed N = 4 SYM theory}},
  \href{http://dx.doi.org/10.1088/1751-8113/47/45/455401}{\emph{J. Phys.}
  {\bfseries A47} (2014) 455401},
  [\href{https://arxiv.org/abs/1308.4420}{{\ttfamily 1308.4420}}].

\bibitem{Sieg:2016vap}
C.~Sieg and M.~Wilhelm, \emph{{On a CFT limit of planar $\gamma_i$-deformed
  $\mathcal{N}=4$ SYM theory}},
  \href{http://dx.doi.org/10.1016/j.physletb.2016.03.004}{\emph{Phys. Lett.}
  {\bfseries B756} (2016) 118--120},
  [\href{https://arxiv.org/abs/1602.05817}{{\ttfamily 1602.05817}}].

\bibitem{Grabner:2017pgm}
D.~Grabner, N.~Gromov, V.~Kazakov and G.~Korchemsky, \emph{{Strongly
  $\gamma$-Deformed $\mathcal{N}=4$ Supersymmetric Yang-Mills Theory as an
  Integrable Conformal Field Theory}},
  \href{http://dx.doi.org/10.1103/PhysRevLett.120.111601}{\emph{Phys. Rev.
  Lett.} {\bfseries 120} (2018) 111601},
  [\href{https://arxiv.org/abs/1711.04786}{{\ttfamily 1711.04786}}].

\bibitem{Zamolodchikov:1980mb}
A.~B. Zamolodchikov, \emph{{`Fishing-net' diagrams as a completely integrable
  system}}, \href{http://dx.doi.org/10.1016/0370-2693(80)90547-X}{\emph{Phys.
  Lett.} {\bfseries 97B} (1980) 63--66}.

\bibitem{Chicherin:2017cns}
D.~Chicherin, V.~Kazakov, F.~Loebbert, D.~Müller and D.-l. Zhong,
  \emph{{Yangian Symmetry for Bi-Scalar Loop Amplitudes}},
  \href{http://dx.doi.org/10.1007/JHEP05(2018)003}{\emph{JHEP} {\bfseries 05}
  (2018) 003}, [\href{https://arxiv.org/abs/1704.01967}{{\ttfamily
  1704.01967}}].

\bibitem{Gromov:2017cja}
N.~Gromov, V.~Kazakov, G.~Korchemsky, S.~Negro and G.~Sizov,
  \emph{{Integrability of Conformal Fishnet Theory}},
  \href{http://dx.doi.org/10.1007/JHEP01(2018)095}{\emph{JHEP} {\bfseries 01}
  (2018) 095}, [\href{https://arxiv.org/abs/1706.04167}{{\ttfamily
  1706.04167}}].

\bibitem{Chicherin:2017frs}
D.~Chicherin, V.~Kazakov, F.~Loebbert, D.~Müller and D.-l. Zhong,
  \emph{{Yangian Symmetry for Fishnet Feynman Graphs}},
  \href{http://dx.doi.org/10.1103/PhysRevD.96.121901}{\emph{Phys. Rev.}
  {\bfseries D96} (2017) 121901},
  [\href{https://arxiv.org/abs/1708.00007}{{\ttfamily 1708.00007}}].

\bibitem{Kazakov:2018hrh}
V.~Kazakov, \emph{{Quantum Spectral Curve of $\gamma$-twisted ${\cal N}=4$ SYM
  theory and fishnet CFT}}, \href{http://dx.doi.org/10.1142/9789813233867_0016,
  10.1142/S0129055X1840010X}{\emph{Rev. Math. Phys.} {\bfseries 30} (2018)
  1840010}, [\href{https://arxiv.org/abs/1802.02160}{{\ttfamily 1802.02160}}].

\bibitem{Gromov:2018hut}
N.~Gromov, V.~Kazakov and G.~Korchemsky, \emph{{Exact Correlation Functions in
  Conformal Fishnet Theory}},
  \href{http://dx.doi.org/10.1007/JHEP08(2019)123}{\emph{JHEP} {\bfseries 08}
  (2019) 123}, [\href{https://arxiv.org/abs/1808.02688}{{\ttfamily
  1808.02688}}].

\bibitem{Korchemsky:2018hnb}
G.~P. Korchemsky, \emph{{Exact scattering amplitudes in conformal fishnet
  theory}}, \href{http://dx.doi.org/10.1007/JHEP08(2019)028}{\emph{JHEP}
  {\bfseries 08} (2019) 028},
  [\href{https://arxiv.org/abs/1812.06997}{{\ttfamily 1812.06997}}].

\bibitem{Basso:2018cvy}
B.~Basso, J.~Caetano and T.~Fleury, \emph{{Hexagons and Correlators in the
  Fishnet Theory}},
  \href{http://dx.doi.org/10.1007/JHEP11(2019)172}{\emph{JHEP} {\bfseries 11}
  (2019) 172}, [\href{https://arxiv.org/abs/1812.09794}{{\ttfamily
  1812.09794}}].

\bibitem{Chowdhury:2019hns}
S.~Dutta~Chowdhury, P.~Haldar and K.~Sen, \emph{{On the Regge limit of Fishnet
  correlators}}, \href{http://dx.doi.org/10.1007/JHEP10(2019)249}{\emph{JHEP}
  {\bfseries 10} (2019) 249},
  [\href{https://arxiv.org/abs/1908.01123}{{\ttfamily 1908.01123}}].

\bibitem{Basso:2018agi}
B.~Basso and D.-l. Zhong, \emph{{Continuum limit of fishnet graphs and AdS
  sigma model}}, \href{http://dx.doi.org/10.1007/JHEP01(2019)002}{\emph{JHEP}
  {\bfseries 01} (2019) 002},
  [\href{https://arxiv.org/abs/1806.04105}{{\ttfamily 1806.04105}}].

\bibitem{Gromov:2019aku}
N.~Gromov and A.~Sever, \emph{{The Holographic Fishchain}},
  \href{http://dx.doi.org/10.1103/PhysRevLett.123.081602}{\emph{Phys. Rev.
  Lett.} {\bfseries 123} (2019) 081602},
  [\href{https://arxiv.org/abs/1903.10508}{{\ttfamily 1903.10508}}].

\bibitem{Gromov:2019bsj}
N.~Gromov and A.~Sever, \emph{{Quantum fishchain in AdS$_{5}$}},
  \href{http://dx.doi.org/10.1007/JHEP10(2019)085}{\emph{JHEP} {\bfseries 10}
  (2019) 085}, [\href{https://arxiv.org/abs/1907.01001}{{\ttfamily
  1907.01001}}].

\bibitem{Gromov:2019jfh}
N.~Gromov and A.~Sever, \emph{{The Holographic Dual of Strongly
  $\gamma$-deformed N=4 SYM Theory: Derivation, Generalization, Integrability
  and Discrete Reparametrization Symmetry}},
  \href{https://arxiv.org/abs/1908.10379}{{\ttfamily 1908.10379}}.

\bibitem{Karananas:2019aab}
G.~K. Karananas, V.~Kazakov and M.~Shaposhnikov, \emph{{Spontaneous Conformal
  Symmetry Breaking in Fishnet CFT}},
  \href{https://arxiv.org/abs/1908.04302}{{\ttfamily 1908.04302}}.

\bibitem{Kazakov:2018qez}
V.~Kazakov and E.~Olivucci, \emph{{Biscalar Integrable Conformal Field Theories
  in Any Dimension}},
  \href{http://dx.doi.org/10.1103/PhysRevLett.121.131601}{\emph{Phys. Rev.
  Lett.} {\bfseries 121} (2018) 131601},
  [\href{https://arxiv.org/abs/1801.09844}{{\ttfamily 1801.09844}}].

\bibitem{Derkachov:2018rot}
S.~Derkachov, V.~Kazakov and E.~Olivucci, \emph{{Basso-Dixon Correlators in
  Two-Dimensional Fishnet CFT}},
  \href{http://dx.doi.org/10.1007/JHEP04(2019)032}{\emph{JHEP} {\bfseries 04}
  (2019) 032}, [\href{https://arxiv.org/abs/1811.10623}{{\ttfamily
  1811.10623}}].

\bibitem{Pittelli:2019ceq}
A.~Pittelli and M.~Preti, \emph{{Integrable fishnet from $\gamma$-deformed
  $\mathcal{N}=2$ quivers}},
  \href{http://dx.doi.org/10.1016/j.physletb.2019.134971}{\emph{Phys. Lett.}
  {\bfseries B798} (2019) 134971},
  [\href{https://arxiv.org/abs/1906.03680}{{\ttfamily 1906.03680}}].

\bibitem{Basso:2017jwq}
B.~Basso and L.~J. Dixon, \emph{{Gluing Ladder Feynman Diagrams into
  Fishnets}},
  \href{http://dx.doi.org/10.1103/PhysRevLett.119.071601}{\emph{Phys. Rev.
  Lett.} {\bfseries 119} (2017) 071601},
  [\href{https://arxiv.org/abs/1705.03545}{{\ttfamily 1705.03545}}].

\bibitem{Bourjaily:2018ycu}
J.~L. Bourjaily, Y.-H. He, A.~J. Mcleod, M.~Von~Hippel and M.~Wilhelm,
  \emph{{Traintracks through Calabi-Yau Manifolds: Scattering Amplitudes beyond
  Elliptic Polylogarithms}},
  \href{http://dx.doi.org/10.1103/PhysRevLett.121.071603}{\emph{Phys. Rev.
  Lett.} {\bfseries 121} (2018) 071603},
  [\href{https://arxiv.org/abs/1805.09326}{{\ttfamily 1805.09326}}].

\bibitem{Penrose:1967wn}
R.~Penrose, \emph{{Twistor algebra}},
  \href{http://dx.doi.org/10.1063/1.1705200}{\emph{J. Math. Phys.} {\bfseries
  8} (1967) 345}.

\bibitem{Penrose:1972ia}
R.~Penrose and M.~A.~H. MacCallum, \emph{{Twistor theory: An Approach to the
  quantization of fields and space-time}},
  \href{http://dx.doi.org/10.1016/0370-1573(73)90008-2}{\emph{Phys. Rept.}
  {\bfseries 6} (1972) 241--316}.

\bibitem{Ward:1990vs}
R.~S. Ward and R.~O. Wells, \emph{{Twistor geometry and field theory}}.
\newblock Cambridge Monographs on Mathematical Physics. Cambridge University
  Press, 1991,
  \href{http://dx.doi.org/10.1017/CBO9780511524493}{10.1017/CBO9780511524493}.

\bibitem{Adamo:2017qyl}
T.~Adamo, \emph{{Lectures on twistor theory}},
  \href{http://dx.doi.org/10.22323/1.323.0003}{\emph{PoS} {\bfseries
  Modave2017} (2018) 003}, [\href{https://arxiv.org/abs/1712.02196}{{\ttfamily
  1712.02196}}].

\bibitem{Witten:2003nn}
E.~Witten, \emph{{Perturbative gauge theory as a string theory in twistor
  space}}, \href{http://dx.doi.org/10.1007/s00220-004-1187-3}{\emph{Commun.
  Math. Phys.} {\bfseries 252} (2004) 189--258},
  [\href{https://arxiv.org/abs/hep-th/0312171}{{\ttfamily hep-th/0312171}}].

\bibitem{Roiban:2004yf}
R.~Roiban, M.~Spradlin and A.~Volovich, \emph{{On the tree level S matrix of
  Yang-Mills theory}},
  \href{http://dx.doi.org/10.1103/PhysRevD.70.026009}{\emph{Phys. Rev.}
  {\bfseries D70} (2004) 026009},
  [\href{https://arxiv.org/abs/hep-th/0403190}{{\ttfamily hep-th/0403190}}].

\bibitem{Mason:2005zm}
L.~J. Mason, \emph{{Twistor actions for non-self-dual fields: A Derivation of
  twistor-string theory}},
  \href{http://dx.doi.org/10.1088/1126-6708/2005/10/009}{\emph{JHEP} {\bfseries
  10} (2005) 009}, [\href{https://arxiv.org/abs/hep-th/0507269}{{\ttfamily
  hep-th/0507269}}].

\bibitem{Boels:2006ir}
R.~Boels, L.~J. Mason and D.~Skinner, \emph{{Supersymmetric Gauge Theories in
  Twistor Space}},
  \href{http://dx.doi.org/10.1088/1126-6708/2007/02/014}{\emph{JHEP} {\bfseries
  02} (2007) 014}, [\href{https://arxiv.org/abs/hep-th/0604040}{{\ttfamily
  hep-th/0604040}}].

\bibitem{Adamo:2011pv}
T.~Adamo, M.~Bullimore, L.~Mason and D.~Skinner, \emph{{Scattering Amplitudes
  and Wilson Loops in Twistor Space}},
  \href{http://dx.doi.org/10.1088/1751-8113/44/45/454008}{\emph{J. Phys.}
  {\bfseries A44} (2011) 454008},
  [\href{https://arxiv.org/abs/1104.2890}{{\ttfamily 1104.2890}}].

\bibitem{Sokatchev:1995nj}
E.~Sokatchev, \emph{{An Action for N=4 supersymmetric selfdual Yang-Mills
  theory}}, \href{http://dx.doi.org/10.1103/PhysRevD.53.2062}{\emph{Phys. Rev.}
  {\bfseries D53} (1996) 2062--2070},
  [\href{https://arxiv.org/abs/hep-th/9509099}{{\ttfamily hep-th/9509099}}].

\bibitem{Chicherin:2016soh}
D.~Chicherin and E.~Sokatchev, \emph{{Demystifying the twistor construction of
  composite operators in ${\mathcal N}=4$ super-Yang–Mills theory}},
  \href{http://dx.doi.org/10.1088/1751-8121/aa6b95}{\emph{J. Phys.} {\bfseries
  A50} (2017) 205402}, [\href{https://arxiv.org/abs/1603.08478}{{\ttfamily
  1603.08478}}].

\bibitem{Cachazo:2004kj}
F.~Cachazo, P.~Svrcek and E.~Witten, \emph{{MHV vertices and tree amplitudes in
  gauge theory}},
  \href{http://dx.doi.org/10.1088/1126-6708/2004/09/006}{\emph{JHEP} {\bfseries
  09} (2004) 006}, [\href{https://arxiv.org/abs/hep-th/0403047}{{\ttfamily
  hep-th/0403047}}].

\bibitem{Boels:2007qn}
R.~Boels, L.~J. Mason and D.~Skinner, \emph{{From twistor actions to MHV
  diagrams}},
  \href{http://dx.doi.org/10.1016/j.physletb.2007.02.058}{\emph{Phys. Lett.}
  {\bfseries B648} (2007) 90--96},
  [\href{https://arxiv.org/abs/hep-th/0702035}{{\ttfamily hep-th/0702035}}].

\bibitem{Bullimore:2010pj}
M.~Bullimore, L.~J. Mason and D.~Skinner, \emph{{MHV Diagrams in Momentum
  Twistor Space}}, \href{http://dx.doi.org/10.1007/JHEP12(2010)032}{\emph{JHEP}
  {\bfseries 12} (2010) 032},
  [\href{https://arxiv.org/abs/1009.1854}{{\ttfamily 1009.1854}}].

\bibitem{Adamo:2011cb}
T.~Adamo and L.~Mason, \emph{{MHV diagrams in twistor space and the twistor
  action}}, \href{http://dx.doi.org/10.1103/PhysRevD.86.065019}{\emph{Phys.
  Rev.} {\bfseries D86} (2012) 065019},
  [\href{https://arxiv.org/abs/1103.1352}{{\ttfamily 1103.1352}}].

\bibitem{Chicherin:2014uca}
D.~Chicherin, R.~Doobary, B.~Eden, P.~Heslop, G.~P. Korchemsky, L.~Mason
  et~al., \emph{{Correlation functions of the chiral stress-tensor multiplet in
  $ \mathcal{N}=4 $ SYM}},
  \href{http://dx.doi.org/10.1007/JHEP06(2015)198}{\emph{JHEP} {\bfseries 06}
  (2015) 198}, [\href{https://arxiv.org/abs/1412.8718}{{\ttfamily 1412.8718}}].

\bibitem{Chicherin:2016fac}
D.~Chicherin and E.~Sokatchev, \emph{{$ \mathcal{N} $ = 4 super-Yang-Mills in
  LHC superspace part I: classical and quantum theory}},
  \href{http://dx.doi.org/10.1007/JHEP02(2017)062}{\emph{JHEP} {\bfseries 02}
  (2017) 062}, [\href{https://arxiv.org/abs/1601.06803}{{\ttfamily
  1601.06803}}].

\bibitem{Chicherin:2016fbj}
D.~Chicherin and E.~Sokatchev, \emph{{$ \mathcal{N} $ = 4 super-Yang-Mills in
  LHC superspace part II: non-chiral correlation functions of the stress-tensor
  multiplet}}, \href{http://dx.doi.org/10.1007/JHEP03(2017)048}{\emph{JHEP}
  {\bfseries 03} (2017) 048},
  [\href{https://arxiv.org/abs/1601.06804}{{\ttfamily 1601.06804}}].

\bibitem{Koster:2016ebi}
L.~Koster, V.~Mitev, M.~Staudacher and M.~Wilhelm, \emph{{Composite Operators
  in the Twistor Formulation of N=4 Supersymmetric Yang-Mills Theory}},
  \href{http://dx.doi.org/10.1103/PhysRevLett.117.011601}{\emph{Phys. Rev.
  Lett.} {\bfseries 117} (2016) 011601},
  [\href{https://arxiv.org/abs/1603.04471}{{\ttfamily 1603.04471}}].

\bibitem{Koster:2016loo}
L.~Koster, V.~Mitev, M.~Staudacher and M.~Wilhelm, \emph{{All tree-level MHV
  form factors in $ \mathcal{N} $ = 4 SYM from twistor space}},
  \href{http://dx.doi.org/10.1007/JHEP06(2016)162}{\emph{JHEP} {\bfseries 06}
  (2016) 162}, [\href{https://arxiv.org/abs/1604.00012}{{\ttfamily
  1604.00012}}].

\bibitem{Chicherin:2016qsf}
D.~Chicherin and E.~Sokatchev, \emph{{Composite operators and form factors in
  ${\mathcal N} = 4$ SYM}},
  \href{http://dx.doi.org/10.1088/1751-8121/aa72fe}{\emph{J. Phys.} {\bfseries
  A50} (2017) 275402}, [\href{https://arxiv.org/abs/1605.01386}{{\ttfamily
  1605.01386}}].

\bibitem{Koster:2016fna}
L.~Koster, V.~Mitev, M.~Staudacher and M.~Wilhelm, \emph{{On Form Factors and
  Correlation Functions in Twistor Space}},
  \href{http://dx.doi.org/10.1007/JHEP03(2017)131}{\emph{JHEP} {\bfseries 03}
  (2017) 131}, [\href{https://arxiv.org/abs/1611.08599}{{\ttfamily
  1611.08599}}].

\bibitem{Brandhuber:2016xue}
A.~Brandhuber, E.~Hughes, R.~Panerai, B.~Spence and G.~Travaglini, \emph{{The
  connected prescription for form factors in twistor space}},
  \href{http://dx.doi.org/10.1007/JHEP11(2016)143}{\emph{JHEP} {\bfseries 11}
  (2016) 143}, [\href{https://arxiv.org/abs/1608.03277}{{\ttfamily
  1608.03277}}].

\bibitem{Mason:2009qx}
L.~J. Mason and D.~Skinner, \emph{{Dual Superconformal Invariance, Momentum
  Twistors and Grassmannians}},
  \href{http://dx.doi.org/10.1088/1126-6708/2009/11/045}{\emph{JHEP} {\bfseries
  11} (2009) 045}, [\href{https://arxiv.org/abs/0909.0250}{{\ttfamily
  0909.0250}}].

\bibitem{Bullimore:2011kg}
M.~Bullimore and D.~Skinner, \emph{{Descent Equations for Superamplitudes}},
  \href{https://arxiv.org/abs/1112.1056}{{\ttfamily 1112.1056}}.

\bibitem{ArkaniHamed:2010kv}
N.~Arkani-Hamed, J.~L. Bourjaily, F.~Cachazo, S.~Caron-Huot and J.~Trnka,
  \emph{{The All-Loop Integrand For Scattering Amplitudes in Planar N=4 SYM}},
  \href{http://dx.doi.org/10.1007/JHEP01(2011)041}{\emph{JHEP} {\bfseries 01}
  (2011) 041}, [\href{https://arxiv.org/abs/1008.2958}{{\ttfamily 1008.2958}}].

\bibitem{Mason:2010yk}
L.~J. Mason and D.~Skinner, \emph{{The Complete Planar S-matrix of N=4 SYM as a
  Wilson Loop in Twistor Space}},
  \href{http://dx.doi.org/10.1007/JHEP12(2010)018}{\emph{JHEP} {\bfseries 12}
  (2010) 018}, [\href{https://arxiv.org/abs/1009.2225}{{\ttfamily 1009.2225}}].

\bibitem{Bullimore:2011ni}
M.~Bullimore and D.~Skinner, \emph{{Holomorphic Linking, Loop Equations and
  Scattering Amplitudes in Twistor Space}},
  \href{https://arxiv.org/abs/1101.1329}{{\ttfamily 1101.1329}}.

\bibitem{Adamo:2011dq}
T.~Adamo, M.~Bullimore, L.~Mason and D.~Skinner, \emph{{A Proof of the
  Supersymmetric Correlation Function / Wilson Loop Correspondence}},
  \href{http://dx.doi.org/10.1007/JHEP08(2011)076}{\emph{JHEP} {\bfseries 08}
  (2011) 076}, [\href{https://arxiv.org/abs/1103.4119}{{\ttfamily 1103.4119}}].

\bibitem{Adamo:2011cd}
T.~Adamo, \emph{{Correlation functions, null polygonal Wilson loops, and local
  operators}}, \href{http://dx.doi.org/10.1007/JHEP12(2011)006}{\emph{JHEP}
  {\bfseries 12} (2011) 006},
  [\href{https://arxiv.org/abs/1110.3925}{{\ttfamily 1110.3925}}].

\bibitem{Mason:2007ct}
L.~J. Mason and M.~Wolf, \emph{{Twistor Actions for Self-Dual Supergravities}},
  \href{http://dx.doi.org/10.1007/s00220-009-0732-5}{\emph{Commun. Math. Phys.}
  {\bfseries 288} (2009) 97--123},
  [\href{https://arxiv.org/abs/0706.1941}{{\ttfamily 0706.1941}}].

\bibitem{Adamo:2013tja}
T.~Adamo and L.~Mason, \emph{{Conformal and Einstein gravity from twistor
  actions}},
  \href{http://dx.doi.org/10.1088/0264-9381/31/4/045014}{\emph{Class. Quant.
  Grav.} {\bfseries 31} (2014) 045014},
  [\href{https://arxiv.org/abs/1307.5043}{{\ttfamily 1307.5043}}].

\bibitem{Haehnel:2016mlb}
P.~Hähnel and T.~McLoughlin, \emph{{Conformal higher spin theory and twistor
  space actions}}, \href{http://dx.doi.org/10.1088/1751-8121/aa9108}{\emph{J.
  Phys.} {\bfseries A50} (2017) 485401},
  [\href{https://arxiv.org/abs/1604.08209}{{\ttfamily 1604.08209}}].

\bibitem{Adamo:2016ple}
T.~Adamo, P.~Hähnel and T.~McLoughlin, \emph{{Conformal higher spin scattering
  amplitudes from twistor space}},
  \href{http://dx.doi.org/10.1007/JHEP04(2017)021}{\emph{JHEP} {\bfseries 04}
  (2017) 021}, [\href{https://arxiv.org/abs/1611.06200}{{\ttfamily
  1611.06200}}].

\bibitem{Adamo:2017xaf}
T.~Adamo, D.~Skinner and J.~Williams, \emph{{Minitwistors and 3d
  Yang-Mills-Higgs theory}},
  \href{http://dx.doi.org/10.1063/1.5030417}{\emph{J. Math. Phys.} {\bfseries
  59} (2018) 122301}, [\href{https://arxiv.org/abs/1712.09604}{{\ttfamily
  1712.09604}}].

\bibitem{Adamo:2013cra}
T.~Adamo, \emph{{Twistor actions for gauge theory and gravity}}, Ph.D. thesis,
  University of Oxford, 2013.
\newblock \href{https://arxiv.org/abs/1308.2820}{{\ttfamily 1308.2820}}.

\bibitem{Lipstein:2013xra}
A.~E. Lipstein and L.~Mason, \emph{{From $d$ logs to dilogs the super
  Yang-Mills MHV amplitude revisited}},
  \href{http://dx.doi.org/10.1007/JHEP01(2014)169}{\emph{JHEP} {\bfseries 01}
  (2014) 169}, [\href{https://arxiv.org/abs/1307.1443}{{\ttfamily 1307.1443}}].

\bibitem{Koster:2014fva}
L.~Koster, V.~Mitev and M.~Staudacher, \emph{{A Twistorial Approach to
  Integrability in $\mathcal N=$ 4 SYM}},
  \href{http://dx.doi.org/10.1002/prop.201400085}{\emph{Fortsch. Phys.}
  {\bfseries 63} (2015) 142--147},
  [\href{https://arxiv.org/abs/1410.6310}{{\ttfamily 1410.6310}}].

\bibitem{Penrose:1969ae}
R.~Penrose, \emph{{Solutions of the zero-rest-mass equations}},
  \href{http://dx.doi.org/10.1063/1.1664756}{\emph{J. Math. Phys.} {\bfseries
  10} (1969) 38--39}.

\bibitem{Eastwood:1981jy}
M.~G. Eastwood, R.~Penrose and R.~O. Wells, \emph{{Cohomology and Massless
  Fields}}, \href{http://dx.doi.org/10.1007/BF01942327}{\emph{Commun. Math.
  Phys.} {\bfseries 78} (1981) 305--351}.

\bibitem{Ward:1977ta}
R.~S. Ward, \emph{{On Selfdual gauge fields}},
  \href{http://dx.doi.org/10.1016/0375-9601(77)90842-8}{\emph{Phys. Lett.}
  {\bfseries A61} (1977) 81--82}.

\bibitem{Woodhouse:1985id}
N.~M.~J. Woodhouse, \emph{{Real methods in twistor theory}},
  \href{http://dx.doi.org/10.1088/0264-9381/2/3/006}{\emph{Class. Quant. Grav.}
  {\bfseries 2} (1985) 257--291}.

\bibitem{Chalmers:1996rq}
G.~Chalmers and W.~Siegel, \emph{{The Selfdual sector of QCD amplitudes}},
  \href{http://dx.doi.org/10.1103/PhysRevD.54.7628}{\emph{Phys. Rev.}
  {\bfseries D54} (1996) 7628--7633},
  [\href{https://arxiv.org/abs/hep-th/9606061}{{\ttfamily hep-th/9606061}}].

\bibitem{Koster:2017fvf}
L.~Koster, \emph{{Form factors and correlation functions in N = 4 super
  Yang-Mills theory from twistor space}}, Ph.D. thesis, Humboldt Universität,
  Berlin, 2017.
\newblock \href{https://arxiv.org/abs/1712.07566}{{\ttfamily 1712.07566}}.

\bibitem{Kulaxizi:2004pa}
M.~Kulaxizi and K.~Zoubos, \emph{{Marginal deformations of N=4 SYM from
  open/closed twistor strings}},
  \href{http://dx.doi.org/10.1016/j.nuclphysb.2006.01.018}{\emph{Nucl. Phys.}
  {\bfseries B738} (2006) 317--349},
  [\href{https://arxiv.org/abs/hep-th/0410122}{{\ttfamily hep-th/0410122}}].

\bibitem{Gao:2006mw}
P.~Gao and J.-B. Wu, \emph{{(Non)-supersymmetric marginal deformations from
  twistor string theory}},
  \href{http://dx.doi.org/10.1016/j.nuclphysb.2008.01.027}{\emph{Nucl. Phys.}
  {\bfseries B798} (2008) 184--197},
  [\href{https://arxiv.org/abs/hep-th/0611128}{{\ttfamily hep-th/0611128}}].

\bibitem{Eastwood:1981b}
M.~G. Eastwood and M.~L. Ginsberg, \emph{{Duality in twistor theory}},
  \href{http://dx.doi.org/10.1215/S0012-7094-81-04812-2}{\emph{Duke Math.J.}
  {\bfseries 48} (1981) 177--196}.

\bibitem{Brandhuber:2004yw}
A.~Brandhuber, B.~J. Spence and G.~Travaglini, \emph{{One-loop gauge theory
  amplitudes in N=4 super Yang-Mills from MHV vertices}},
  \href{http://dx.doi.org/10.1016/j.nuclphysb.2004.11.023}{\emph{Nucl. Phys.}
  {\bfseries B706} (2005) 150--180},
  [\href{https://arxiv.org/abs/hep-th/0407214}{{\ttfamily hep-th/0407214}}].

\bibitem{Bena:2004xu}
I.~Bena, Z.~Bern, D.~A. Kosower and R.~Roiban, \emph{{Loops in twistor space}},
  \href{http://dx.doi.org/10.1103/PhysRevD.71.106010}{\emph{Phys. Rev.}
  {\bfseries D71} (2005) 106010},
  [\href{https://arxiv.org/abs/hep-th/0410054}{{\ttfamily hep-th/0410054}}].

\bibitem{Chicherin:2017bxc}
D.~Chicherin and E.~Sokatchev, \emph{{Conformal anomaly of generalized form
  factors and finite loop integrals}},
  \href{http://dx.doi.org/10.1007/JHEP04(2018)082}{\emph{JHEP} {\bfseries 04}
  (2018) 082}, [\href{https://arxiv.org/abs/1709.03511}{{\ttfamily
  1709.03511}}].

\bibitem{Usyukina:1992jd}
N.~I. Usyukina and A.~I. Davydychev, \emph{{An Approach to the evaluation of
  three and four point ladder diagrams}},
  \href{http://dx.doi.org/10.1016/0370-2693(93)91834-A}{\emph{Phys. Lett.}
  {\bfseries B298} (1993) 363--370}.

\end{thebibliography}\endgroup
\bibliographystyle{JHEP}

\end{document}